\documentclass[journal,twocolumn]{IEEEtran}

\usepackage[cmex10]{amsmath}
\usepackage{amssymb}
\usepackage{cite}
\usepackage{url}
\usepackage{graphicx}
\usepackage{array,color}
\usepackage{multirow}
\usepackage{amsmath}
\usepackage{stfloats}
\usepackage{graphicx}
\usepackage{subfigure}
\usepackage{tabularx}
\usepackage{epsfig,color,balance,cite}
\usepackage{setspace}
\usepackage{bm}
\usepackage{textcomp}
\usepackage{algorithmic}
\usepackage{algorithm}
\usepackage{comment}
\usepackage{subfigure}
\usepackage{caption}
\usepackage{graphicx}
\usepackage{setspace}
\usepackage{diagbox}
\usepackage{float}
\usepackage{epstopdf}
\usepackage{caption}
\UseRawInputEncoding

\usepackage{subfigure}

\usepackage{caption}
\usepackage{graphicx}
\usepackage{setspace}
\usepackage{diagbox}

\usepackage{multirow}
\usepackage{graphicx}
\makeatletter

\newcommand{\Rmnum}[1]{\expandafter\@slowromancap\romannumeral #1@}
\makeatother
\usepackage{booktabs}
\usepackage{amsmath}

\makeatother
\pdfoutput=1

\begin{document}

\title{Channel Estimation for mmWave MIMO-OFDM Systems in High-Mobility Scenarios:  Instantaneous Model or Statistical Model?}

\author{Ruizhe Wang, Hong Ren, Cunhua Pan, Gui Zhou, Ruisong Weng and Jiangzhou Wang, \IEEEmembership{Fellow, IEEE}%
\thanks{R. Wang, H. Ren, C. Pan and R. Weng are with National Mobile Communications Research Laboratory, Southeast University, Nanjing, China. (e-mail: {rzwang, hren, cpan, ruisong\_weng}@seu.edu.cn).
G. Zhou is with the Institute for Digital Communications, Friedrich-Alexander-University Erlangen-N¨urnberg (FAU), 91054 Erlangen, Germany (email: gui.zhou@fau.de).
Jiangzhou Wang is with the School of Engineering, University of Kent, CT2 7NZ Canterbury, U.K. (e-mail: j.z.wang@kent.ac.uk).
\emph{Corresponding authors: Hong Ren and Cunhua Pan}}
}
\maketitle


\begin{abstract}
Classical linear statistical models, like the first-order auto-regressive (AR) model, are commonly used as channel model in high-mobility scenarios. However, compared to sub-6G, the effect of Doppler frequency shifts is more significant at millimeter wave (mmWave) frequencies, and the effectiveness of the statistical channel model in high-mobility mmWave scenarios should be reconsidered. In this paper, we investigate the channel estimation for mmWave multiple-input multiple-output-(MIMO) orthogonal frequency division multiplexing (OFDM) systems in high-mobility scenarios, with the focus on the comparison between the instantaneous channel model and the statistical channel model. For the instantaneous model, by leveraging the low-rank nature of mmWave channels and the multidimensional characteristics of MIMO-OFDM signals across space, time, and frequency, the received signals are structured as a fourth-order tensor fitting a low-rank CANDECOMP/PARAFAC (CP) model. Then, to solve the CP decomposition problem, an estimation of signal parameters via rotational invariance techniques (ESPRIT)-type decomposition based channel estimation method is proposed by exploring the Vandermonde structure of factor matrix, and the channel parameters are then estimated from the factor matrices. We analyze the uniqueness condition of the CP decomposition and develop a concise derivation of the Cramer-Rao bound (CRB) for channel parameters. Simulations show that our method outperforms the existing benchmarks. Furthermore, the results based on the wireless environment generated by Wireless InSite verify that the channel estimation based on the instantaneous channel model performs better than that based on the statistical channel model. Therefore, the instantaneous channel model is recommended for designing channel estimation algorithm for mmWave systems in high-mobility scenarios.

\end{abstract}


\IEEEpeerreviewmaketitle
%
\section{Introduction}
Millimeter wave (mmWave) communication has been considered as one of the key technologies in fifth-generation (5G) and future sixth-generation (6G) wireless communication systems, which leverages the huge bandwidth at mmWave frequency to support high throughput \cite{mmWave}. Additionally, massive multiple-input multiple-output (MIMO) utilizes the degrees of freedom provided by its large number of antennas to enable the full utilization of spatial resources, thereby improving spectral efficiency \cite{mmwavemMIMO}. The multi-antenna array also helps improve the beamforming precision, signal-to-noise ratio (SNR) and the spectrum efficiency \cite{mmwaveMIMO2}. To reap the benefits promised by mmWave MIMO systems, accurate channel state information (CSI) is required. 

Channel estimation for mmWave MIMO systems has been extensively studied in the existing literature \cite{CS1,CS2,CS3,R1,shijin,R2,R3}. For example, the compressed sensing (CS) method for channel estimation was applied in mmWave MIMO systems \cite{CS1,CS2,CS3} by exploring the sparse property of mmWave channel. In \cite{R1,R2,shijin,R3}, the frequency-selective MIMO channel estimation problem was addressed using tensor decomposition-based methods. To better evaluate the performance of algorithms, the Cramer-Rao bound (CRB) is widely used in channel estimation to describe the performance bound of channel estimation algorithms, providing a benchmark to evaluate the performance of the proposed algorithm \cite{performance1,performance2,zhouzhoujsac,crb1,tuowu}.

It is worth noting that the aforementioned studies predominantly assumed a frequency-selective, time-invariant channel, which did not take into account the high-mobility scenarios. To support the high date rate requirements in high-mobility scenarios, mmWave communication has been applied in highways \cite{highway}, high-speed trains \cite{HST,HST2}, unmanned aerial vehicles (UAVs) \cite{UAV} and other scenarios. Thus, the double-selectivity of mmWave systems in high-mobility scenarios should be considered. 
The statistical linear channel models (i.e., the first order auto-regressive (AR) model \cite{statistical}) were extensively used in the design of time-varying channel estimation algorithm \cite{2024tcomSOMP,CS5,CS6}. 
For example, in  \cite{2024tcomSOMP}, the authors investigated channel estimation for MIMO-orthogonal frequency division multiplexing (OFDM) systems based on Kalman Filter (KF)-CS algorithm. In \cite{CS5,CS6}, the KF-sparse Bayesian learning (SBL) framework was employed in mmWave MIMO-OFDM channel estimation. However, the SBL-based algorithm suffers from high-computational complexity. 
For mmWave systems in high-mobility scenarios, the high Doppler frequency leads to rapid non-linear variations in the instantaneous channel gains, especially the fast phase variations. This raises the question about whether this classical linear statistical channel model is suitable or not for algorithm design in practical scenarios.

Recently, there have been several contributions on channel estimation for mmWave MIMO systems in high-mobility scenarios based on the instantaneous channel model \cite{qinqibotvt,mmBOMP,convergence2,ryzhang}. In \cite{qinqibotvt,mmBOMP}, the channel estimation problem was addressed by OMP-based methods. However, the performance of the parameter estimation is limited due to the on-grid estimation. In \cite{convergence2,ryzhang}, a tensor decomposition based channel estimation method was proposed for MIMO-OFDM systems. Nonetheless, the method relies on the alternating least squares (ALS) algorithm, whose convergence is not guaranteed, especially when the tensor rank exceeds 2 \cite{convergence1}. Furthermore, none of the above studies compared the instantaneous channel model with the statistical channel model in terms of the estimation performance. To the best of our knowledge, the comparison of the performance of channel estimation based on statistical model and the instantaneous model for mmWave massive MIMO-OFDM systems in high-mobility scenarios has not been investigated in the literature.

Motivated by the above discussions, in this paper, we investigate the uplink channel estimation for hybrid-structured mmWave massive MIMO-OFDM systems based on the instantaneous channel model. And we focus on the comparison between the instantaneous channel model and the statistical channel model. 
For the instantaneous channel model, due to the multi-dimensional nature of MIMO-OFDM signals, estimating coupling channel parameters from a complex multilinear array is crucial. Fortunately, tensor theory offers an effective, low-complexity method for array decomposition and feature extraction \cite{SPM,tensor,R4}. We propose a fourth-order tensor model that exploits multi-dimensional property of MIMO-OFDM signals and the low-rank property of the mmWave channel. By utilizing the Vandermonde structure of the factor matrix, the tensor decomposition problem is solved using spatial smoothing and the ESPRIT algorithm, avoiding the non-convergence of the ALS algorithm. Additionally, maximum likelihood (ML) estimation on the resulting fibers improves accuracy over on-grid-searching CS algorithms.
To analyze the parameter estimation performance for the algorithms based on the instantaneous channel model, a more concise derivations of CRB is proposed compared to that in \cite{zhouzhoujsac,crbderive}. For the statistical channel model, the KF-CS algorithm proposed in \cite{2024tcomSOMP} is applied, where the time-varying channel gains are estimated based on Kalman Filtering, and the angles are estimated based on the CS algorithm. Finally, to compare the channel estimation performance of the instantaneous channel model and the statistical channel model, we use Wireless Insites \cite{WI} to generate the practical wireless environment in the simulations. Simulation results show that under the same pilot scheme, the algorithm based on the instantaneous channel model performs better than that based on the statistical channel model. The main contributions of this paper are summarized as follows:
\begin{itemize}
	\item {\textit{\textbf {Novel Mathematical Characterization:}}} We investigate a MIMO-OFDM system in which the base station (BS) communicates with a high-speed mobile station (MS) based on the instantaneous channel model.  By leveraging the low-rank property of mmWave channel and multi-dimensional property of space, time and frequency domains in MIMO-OFDM signals, the received signals are structured into a fourth-order tensor, which fits a CP model.  
	\item {\textit{\textbf {Decomposition based feature extraction:}}} By exploring the low-rank property of mmWave channel and the Vandermonde structure of the time-domain factor matrix, an ESPRIT-type decomposition based algorithm is proposed to complete the fitting of the CP model. Based on the decomposed four factor matrices,  the channel parameters are estimated, including angles, delays, channel gains, and Doppler shifts. 
	\item {\textit{\textbf {Theoretical Analysis:}}} The uniqueness condition of the CP decomposition for the proposed fourth-order tensor is analyzed. Furthermore, to gain deeper insights in performance analysis, the CRB for the channel parameters is derived to compare the estimation performance of the proposed method with that of benchmarks. We propose a more concise derivation of the CRB by utilizing tensor vectorization and the property of the Kronecker product. The proposed derivation prevents the complex element indexing when calculating the fisher matrix (FIM) \cite{zhouzhoujsac,crbderive}.
	\item {\textit{\textbf {Effectiveness Validation:}}} Simulation results demonstrate the superior performance of the proposed method over other benchmarks based on the instantaneous channel model. Furthermore, the channel estimation methods are evaluated in a more realistic simulation environment, where the channel data are generated by the practical channel generation software of Wireless InSites \cite{WI}, with the focus on the comparison between the algorithms based on the instantaneous channel model and that based on the statistical channel model. The results verify that the algorithms based on the instantaneous channel model are more effective. Therefore, the instantaneous channel model is recommended for designing channel estimation algorithm for mmWave systems in high-mobility scenarios. 
\end{itemize}

The rest of this paper is organized as follows. Section \ref{notations} introduces the notations and the preliminaries about tensor theory. Section \ref{systemmodel} describes the channel model for the time-varying MIMO-OFDM channel and formulates the channel estimation problem. In Section \ref{channelestimation}, we propose the ESPRIT-type decomposition based channel estimation method, and we also introduce the ALS-based channel estimation method. The uniqueness condition of the CP decomposition method is then analyzed. 
The CRB results for channel parameters are derived in Section \ref{CRB}. In Section \ref{complexity}, the computational complexity of the proposed methods and the benchmarks are analyzed. Simulation results are provided in Section \ref{simulation}. Finally, conclusions are drawn in Section \ref{conclusion}.

\section{Notations and Preliminaries}\label{notations}
\subsection{Notations:}
In this paper, the following notations are used. Lowercase letter, boldface lowercase letter, boldface uppercase letter and the calligraphy letter denote the scalars, vectors, matrices and tensors, respectively, i.e., $y$, ${\bf y}$, ${\bf Y}$, $\mathcal{Y}$. The operations $(\cdot)^{\ast}$, $(\cdot)^{\mathrm T}$, $(\cdot)^{\mathrm H}$, $(\cdot)^{\mathrm{-T}}$, $\left( \cdot \right) ^{\dagger}$, $\|\cdot\|$ and $\|\cdot\|_{F}$ represent the conjugate, transpose,  transpose-inverse, pseudo-inverse, Hermitian (conjugate transpose), 2-norm and Frobenius norm,  respectively. The symbol $\delta \left( \cdot \right) $ denotes the delta function. The operator $\mathrm{unvec}_{M\times N}\left( \cdot \right) $ denotes the operation that reshapes an $MN\times 1$ vector into an $M\times N$ matrix. The notations $\mathbb{E}(\cdot)$ and $\mathbb{C}$ represent the expectation operator and the complex field, respectively. The symbol $a_{ij}$ refers to the $(i,j)$th entry of matrix $\bf A$. The $n\times 1$ all-one vector and $n\times n$ identity matrix are represented by $\mathbf{1}_n$ and ${\bf I}_n$, respectively. The circularly-symmetric complex Gaussian distribution is denoted as $\mathcal{CN}({\bm{\mu}}, {\bf C})$, where ${\bm{\mu}}$ is the mean vector and ${\bf C}$ is the covariance matrix, respectively. The operators $\circ$, $\otimes $ and $\odot $ denote outer product, Kronecker product and Khatri-Rao product, respectively. Denote $\mathrm{D}\left( {\bf a} \right) $, $\mathrm{D}_n\left({\bf A} \right) $ and $\mathrm{d}\left( {\bf A} \right) $ as the diagonal matrix formed by vector $\bf a$, the diagonal matrix formed by the $n$-th row of matrix ${\bf A}$ and the diagonal element vector of matrix ${\bf A}$, respectively. The symbol $\mathcal{I}_{N,R}$ denotes the $R$-th order identity tensor with dimensions $N\times N\times \cdots \times N$.
\vspace{-0.7cm}

\subsection{Tensor Preliminaries}
Some preliminaries on tensor and CP decomposition are provided for better readability. Further details can be found in \cite{tensor,SPM}. 

In this paper, a tensor represents a multidimensional array \cite{tensor}. One way array of tensor is called fiber, which is defined by fixing all the indices constant but one. Two fibers of tensor form a slice, which is defined by fixing all the indices constant but two. An $N$-th order tensor $\mathcal{X} \in \mathbb{C} ^{I_1\times I_2\times \dots \times I_N}$ is a rank-one tensor if it can be expressed as the outer product of $N$ vectors, i.e.
\begin{equation}
	\mathcal{X} =\mathbf{a}^{\left( 1 \right)}\circ \mathbf{a}^{\left( 2 \right)}\circ \cdots \circ \mathbf{a}^{\left( N \right)}, \mathbf{a}^{\left( i \right)}\in \mathbb{C} ^{I_i\times 1}.
\end{equation}
The CP decomposition represents an $N$-th order tensor $\mathcal{X} \in \mathbb{C} ^{I_1\times I_2\times \dots \times I_N}$ as a sum of rank-one tensors, i.e.
\begin{equation}\label{CPDcomp}
	\mathcal{X} =\sum_{r=1}^R{\mathbf{a}_{r}^{\left( 1 \right)}\circ \mathbf{a}_{r}^{\left( 2 \right)}\circ \cdots \circ \mathbf{a}_{r}^{\left( N \right)}},
\end{equation}
where $R$ is the rank of tensor. The factor matrix according to the $n$-th mode is defined as $\mathbf{A}^{\left( n \right)}\triangleq \left[ \mathbf{a}_{1}^{\left( n \right)},\cdots ,\mathbf{a}_{R}^{\left( n \right)} \right] \in \mathbb{C} ^{I_n\times R}$. Then, (\ref{CPDcomp}) can be equivalently represented as
\begin{align}\label{CPDcomp2}
	\mathcal{X}\nonumber &=\mathbf{A}^{\left( 1 \right)}\times _2\mathbf{A}^{\left( 2 \right)}\times _3\cdots \times _N\mathbf{A}^{\left( N \right)}\nonumber\\
	\\
	&=[\kern-0.15em[\mathbf{A}^{\left( 1 \right)},\mathbf{A}^{\left( 2 \right)},\cdots ,\mathbf{A}^{\left( N \right)}]\kern-0.15em],\nonumber
\end{align}
where the operator $\times_n$ represents the mode-$n$ product. The mode-$n$ unfolding of $\mathcal{X}$ is defined as \cite{Nicholas}
\begin{align}
	\mathbf{X}_{\left( n \right)}&=\left( \mathbf{A}^{\left( N \right)}\odot \cdots \mathbf{A}^{\left( n+1 \right)}\odot \mathbf{A}^{\left( n-1 \right)}\odot \cdots \mathbf{A}^{\left( 1 \right)} \right) \mathbf{A}^{\left( n \right) \mathrm{T}}\nonumber
	\\
	&\qquad\in \mathbb{C} ^{\left( \prod_{i\ne n}^N{I_i} \right) \times I_n}.
\end{align}
The vectorization of the mode-$n$ unfolding of the $N$-th order tensor $\mathcal{X}$ is
\begin{align}
	&\mathrm{vec}\left( \mathbf{X}_n \right) \nonumber\\
	&= \mathbf{A}^{\left( n \right)}\odot \left( \mathbf{A}^{\left( N \right)}\odot \cdots \mathbf{A}^{\left( n+1 \right)}\odot \mathbf{A}^{\left( n-1 \right)}\odot \cdots \mathbf{A}^{\left( 1 \right)} \right)\mathbf{1}_R  .
\end{align}

\section{System Model}\label{systemmodel}
\subsection{Channel Model in High-Mobility Scenario}\label{Time Varying Channel Model}
\begin{figure*}[ht]
	\vspace{-0.2cm}
	\begin{minipage}[t]{1\linewidth}
		\centering
		\includegraphics[width=1.0\linewidth]{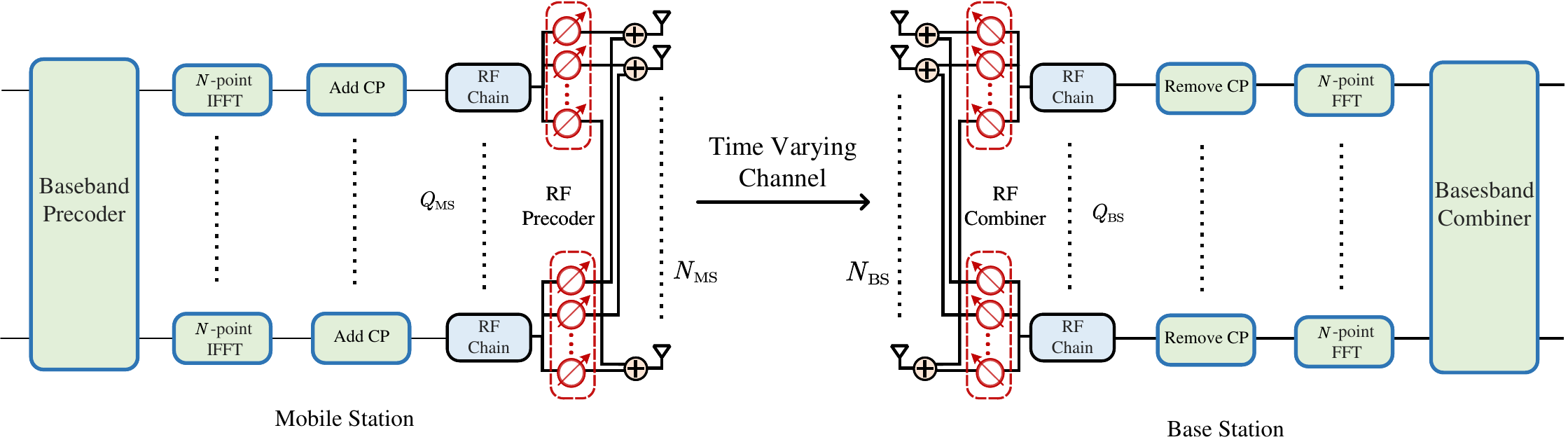}
		\vspace{-0.2cm}
		\caption{mmWave MIMO-OFDM Systems in High-Mobility Scenarios.}
		\label{system}
	\end{minipage}%
	\hfill
	\vspace{-0.3cm}
\end{figure*}

Consider an uplink mmWave massive MIMO-OFDM system shown in Fig.~\ref{system}, which consists of a BS and an MS, where the BS is equipped with $N_{\text{BS}}$ antennas and $Q_{\text{BS}}$ RF chains, and the MS is equipped with $N_{\text{MS}}$ antennas and $Q_{\text{MS}}$ radio frequency (RF) chains. Uniform linear arrays (ULAs) are assumed to be equipped at both the BS and the MS. Without loss of generality, the number of RF chains is less than that of antennas, i.e. $Q_{\text{MS}}<N_{\text{MS}}$ and $Q_{\text{BS}}<N_{\text{BS}}$. The total number of OFDM subcarriers is $N$, in which $K$ subcarriers are selected for channel estimation. In the following sections, we only consider the single user case, and the multiuser case can be simply extended by allocating different bandwith parts (BWPs) to different users and implementing channel estimation on different subcarrier sets. Assuming that there are $L$ propagation paths between the BS and the MS, the channel matrix in time and delay domain is given by
\begin{equation}
\mathbf{H}\left[ t,\tau \right] =\sum_{l=1}^L{\alpha _{t,l}\mathbf{a}_{N_{\mathrm{BS}}}\left( \theta _l \right) \mathbf{a}_{N_{\mathrm{MS}}}^{\mathrm{T}}\left( \phi _l \right) \delta \left( \tau -\tau _l \right)},
\end{equation}
where $t$ and $\tau$ denote the scale in the continuous time domain and delay domain, respectively,  $\alpha_{t,l}$ denotes the complex channel gain following the complex Gaussian distribution, $\theta_{l}$ and $\phi_l$ denote the angle of arrival (AoA) and the angle of departure (AoD), respectively, and $\tau_l$ denotes the propagation delay of the $l$-th path. Moreover, $\mathbf{a}_{N_{\text{BS}}}(\cdot)$ and $\mathbf{a}_{N_{\text{MS}}}(\cdot)$ denote the array steering vectors of the BS and the MS, respectively. For a simple ULA, the array steering vector ${\bf a}_X(x)\in\mathbb{C}^{X\times 1}$ is represented by
\begin{equation}
	\mathbf{a}_X\left( x \right) =\left[ 1,e^{j2\pi \frac{d}{\lambda _{\mathrm{c}}}\cos\mathrm{(}x)},\cdots ,e^{j2\pi (X-1)\frac{d}{\lambda _{\mathrm{c}}}\cos\mathrm{(}x)} \right] ^{\mathrm T},
\end{equation}
where $d$ is the antenna spacing, $\lambda _{\mathrm{c}}$ is the carrier wavelength and $x$ is the angle (AoA for BS and AoD for MS, respectively). It is assumed that the antenna spacing in every type of ULA satisfies $d=\lambda _{\mathrm{c}}/2$. 

In high-mobility scenarios, rapid channel changes occur due to Doppler frequency shift, resulting in a short channel coherence time interval \cite{wcx}. In 5G mmWave MIMO systems, the subcarrier spacing (SCS) can be set larger than in sub-6G systems, leading to shorter symbol duration. 5G New Radio (NR) supports multiple OFDM numerologies with SCS of ${\text 2}^\mu \cdot{\text {15}}$ kHz, where $\mu$ ranges from 0 to 6 \cite{3GPP13810101}. MmWave bands have more bandwidth, allowing larger SCS. Thus, during a mini-slot, the variation in channel gain $\alpha_{t,l}$, angles, and Doppler shift effects are minimal due to the short symbol interval. For instance, with $N_{\mathrm{s}}=7$ OFDM symbols per mini-slot, $\mu=\text{5}$ yields an SCS of 480 kHz and a mini-slot duration of 14.6 µs. Given a speed of 30 m/s and a carrier frequency of 30 GHz, the coherence time is 333.33 µs. Therefore, it is reasonable to assume that the channel remains invariant during a mini-slot, with constant channel parameters (angles, delays, Doppler shifts) across several mini-slots.

Next, we briefly introduce the differences between the statistical channel model and the instantaneous channel model. Specifically, in the statistical channel model, the time-varying channel gain in the $m$-th mini-slot is modeled as a first-order AR model \cite{statistical}
\begin{equation}
	\alpha _{m,l}=\rho \alpha _{m-1,l}+\psi _{m-1,l},
\end{equation}
where $\rho$ denotes the correlation coefficient and $\psi _{m,l}\sim \mathcal{C} \mathcal{N} (0,1-\rho ^2)$ denotes the innovation noise. Based on the statistical model, we have $\rho =J_0(2\pi f_{\max}^{\text{d}}N_{\text s}T_{\text s})$, where $f_{\max}^{\text{d}}$ denotes the maximum Doppler frequency shift, $N_{\text s}$ denotes the number of symbols in one mini-slot, $T_{\text s}$ denotes the symbol duration. In OFDM systems, we have $T_s={{1}/{\Delta f}}$. Based on the statistical model, the discrete time and delay domain channel matrix channel matrix $\mathbf{H}^{\text{stat}}_m\left[ n,\tau \right]$ is given by
\begin{equation}
	\mathbf{H}^{\text{stat}}_m\left[ n,\tau \right] =\sum_{l=1}^L{\alpha _{m,l}\mathbf{a}_{N_{\mathrm{BS}}}\left( \theta _l \right) \mathbf{a}_{N_{\mathrm{MS}}}^{\mathrm{T}}\left( \phi _l \right) \delta \left( \tau -\tau _l \right)}.
\end{equation}

For instantaneous channel model, the discrete time and delay domain channel matrix $\mathbf{H}^{\text{ins}}_m\left[n,\tau \right]$ at the $n$-th symbol in the $m$-th mini-slot is given by
\begin{align}\label{Hm}
	&\mathbf{H}^{\text{ins}}_m\left[n,\tau \right] \nonumber\\ 
	&=\sum_{l=1}^L{\alpha _le^{j2\pi f_{l}^\mathrm{d}\left(\tau _l+\left( m-1 \right) *N_{\text s}*T_{\text s} \right)}\mathbf{a}_{N_{\text{BS}}}\left( \theta _{l} \right) \mathbf{a}_{N_{\text{MS}}}^{\mathrm{T}}\left( \phi _{l} \right) \delta \left( \tau -\tau _l \right)},
\end{align}
where $\alpha_l$ and $f_{l}^\mathrm{d}$ denote the complex channel gain and the Doppler frequency shift of the $l$-th path, respectively. In the following derivations, we only consider the channel estimation based on the instantaneous channel model. Unless otherwise specified, $\mathbf{H}_{m,k}$  in the following text  refers to the instantaneous channel model. For the statistical channel model, we directly apply the KF-CS algorithm in \cite{2024tcomSOMP} for comparison in simulation results.

\vspace{-0.4cm}
\subsection{Signal Model}
Denote $N$ and $K$ as the number of total subcarriers and the number of subcarriers that transmit pilot signals, respectively. The MS transmits the pilot symbols at the $k$-th subcarrier precoded by the digital precoder and the common RF precoder as 
\begin{equation}
	\mathbf{s}_{m,k}[n] =\mathbf{F}_{\mathrm{RF}}\mathbf{F}_m\mathbf{x}_{m,k}[n] ,
\end{equation}
where $\mathbf{x}_{m,k}[n], n=1,2,\cdots,N_{\text s}$ denotes the pilot signal in the $n$-th symbol in the $m$-th mini-slot at the $k$-th subcarrier, $\mathbf{F}_{\text{RF}}$ and $\mathbf{F}_m$ denote the common RF precoder for all subcarriers and the digital precoding matrix in the $m$-th mini-slot. In mmWave MIMO-OFDM systems, the symbol ${\bf x}_{m,k}[n]$ at the $k$-th subcarrier is first precoded by digital precoder, then the $N$-points inverse Discrete Fourier Transform (IDFT) is processed to transfer the frequency domain symbols into time domain signals. Finally, the time domain signals are processed by common RF precoder and transmitted by antennas.

After receiving the time domain signals, the BS first processes the signals by common RF combiner. Then, the cyclic prefix is removed and the $N$-points DFT is processed to obtain the frequency domain symbols \cite{mmwaveofdm}. Given the time and delay domain channel in (\ref{Hm}), the frequency domain channel in the $m$-th mini-slot at the $k$-th subcarrier is given by
\begin{align}
	&\mathbf{H}_{m,k}\nonumber\\ &=\sum_{l=1}^L{\alpha _le^{j2\pi f^\mathrm{d}_{l}\left( \tau _l+\left( m-1 \right) *N_s*T_{\text s} \right)-j\frac{2\pi}{N}k f_{\text s}\tau _l }\mathbf{a}_{N_{\text{BS}}}\left( \theta _{l} \right) \mathbf{a}^{\mathrm T}_{N_{\text{MS}}}\left( \phi _{l} \right)}\nonumber\\
	&=\mathbf{A}_{N_{\text {BS}}}\mathrm{D}_k({\bf C})\mathrm{D}_m({\bf D})\mathbf{A}^{\mathrm T}_{N_{\text {MS}}},
\end{align}
where $f_{\text s}$ denotes the OFDM sampling frequency, and
\begin{align}
	\mathbf{A}_{N_{\text {BS}}}&=\left[ \begin{matrix}
		\mathbf{a}_{N_{\text {BS}}}\left( \theta _{1} \right),		\cdots,		\mathbf{a}_{N_{\text {BS}}}\left( \theta _{L} \right)\\
	\end{matrix} \right]\in \mathbb{C} ^{N_{\text{BS}}\times L}, \\ 
	\mathbf{A}_{N_{\text {MS}}}&=\left[ \begin{matrix}
		\mathbf{a}_{N_{\text {MS}}}\left( \phi _{1} \right),		\cdots,	\mathbf{a}_{N_{\text {MS}}}\left( \phi _{L} \right)\\
	\end{matrix} \right]\in \mathbb{C} ^{N_{\text{MS}}\times L} \label{AMS},
\end{align}
and ${\mathbf{C}}=\left[ \mathbf{c}_1,\cdots ,\mathbf{c}_L \right] \in \mathbb{C} ^{K\times L}$, ${\mathbf{D}}=\left[ \mathbf{d}_1,\cdots ,\mathbf{d}_L \right] \in \mathbb{C} ^{M\times L}$. In high-speed mobile scenarios, the large subcarrier spacing reduces the impact of inter-subcarrier interference (ICI) \cite{highway,HST,HST2}. Therefore, in the following context, ICI will be treated as part of the noise. For convenience, we assume that the $K$ subcarriers are adjacent though they are not necessary adjacent in our method. And we have
\begin{align}
	\mathbf{c}_l&\triangleq\alpha _l\left[ e^{-j\frac{2\pi}{N}\left( f_{\text s}\tau _l \right) \left( 1-\frac{f^\mathrm{d}_{l}}{f_{\text s}}N \right)},\cdots ,e^{-j\frac{2\pi}{N}\left( f_{\text s}\tau _l \right) \left( K-\frac{f^\mathrm{d}_{l}}{f_{\text s}}N \right)} \right]^{\mathrm T} \label{cl}\\
	{\bf d}_l&\triangleq\left[ 1,e^{j2\pi f^\mathrm{d}_{l}N_{\text s}T_{\text s}},\cdots ,e^{j2\pi f^\mathrm{d}_{l}\left( M-1 \right) N_{\text s}T_{\text s}} \right] ^{\mathrm T}. \label{dl}
\end{align}
From (\ref{cl}), it can be found that the $l$-th column of ${\bf C}$ contains the offset $\frac{f^\mathrm{d}_{l}}{f_{\text s}}N$ caused by the Doppler frequency shift of the $l$-th channel path. From (\ref{dl}), the phase shift effect on received signals in each mini-slot caused by the Doppler frequency shift is demonstrated. 

\begin{figure}[ht]
	\vspace{-0.5cm}
	\begin{minipage}[t]{1\linewidth}
		\centering
		\includegraphics[width=1.0\linewidth]{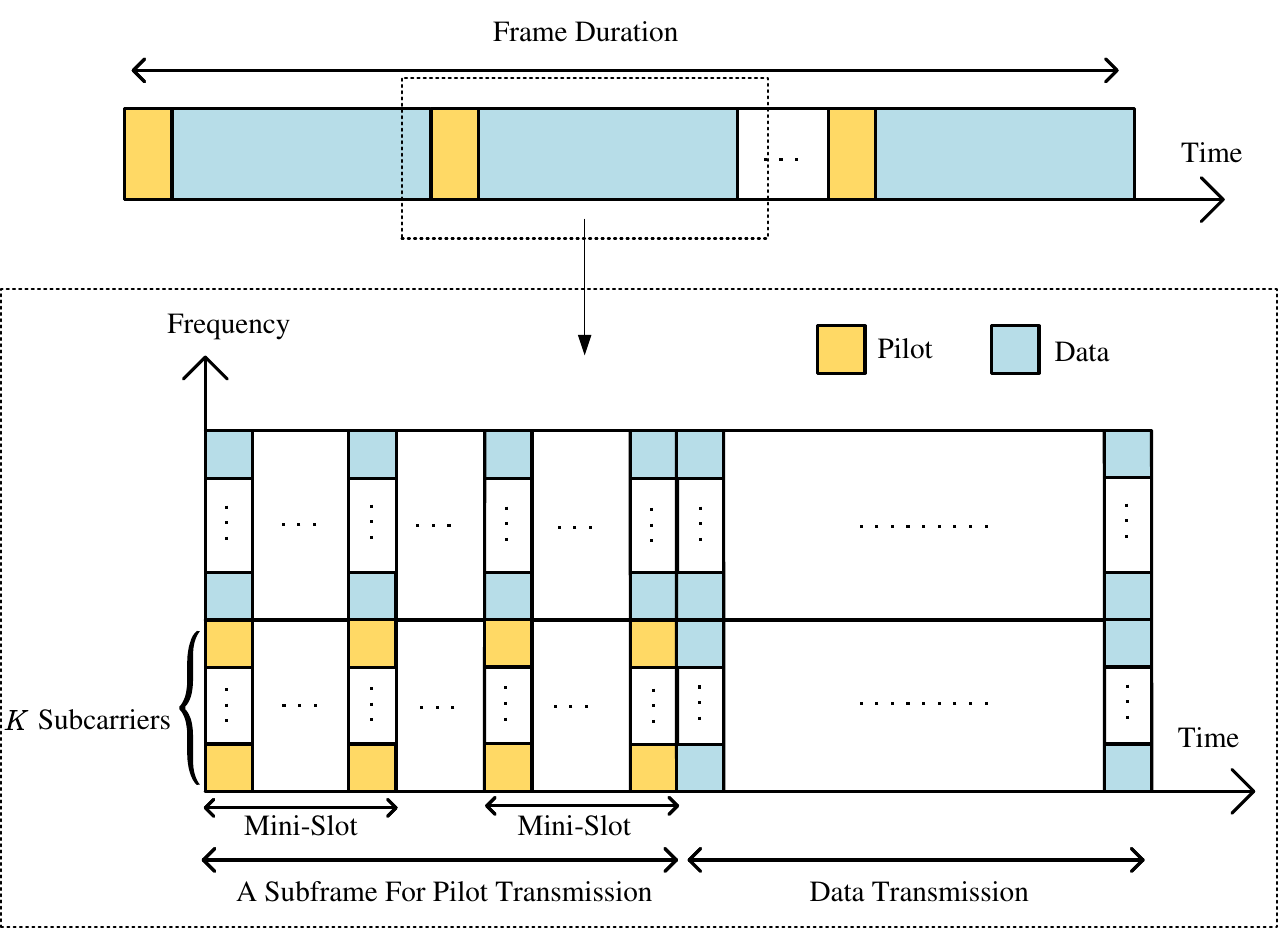}
		\vspace{-0.2cm}
		\caption{The transmission structure for time varying channel estimation in OFDM systems.}
		\label{ofdmpilot}
	\end{minipage}%
	\hfill
	\vspace{-0.7cm}
\end{figure}

Then, we introduce the pilot transmission scheme in mmWave MIMO-OFDM systems in high-mobility scenarios. At each mini-slot, the MS transmits $N_{\text{s}}$ pilot symbols, and the BS receives the pilot signals and employs $Q_{\text{BS}}$ RF combining vectors $\{{\bf w}_q\}$. Specifically, the frame structure and the pilot transmission scheme are shown in Fig.~\ref{ofdmpilot}. The processed signal in the $q$-th RF chain at the $k$-th subcarrier of the $m$-th mini-slot is given by
\begin{equation}
	\mathbf{y}_{q,m,k}=\mathbf{w}_{q}^{\mathrm T}\left(\mathbf{H}_{m,k} \mathbf{S}_{m,k}+\mathbf{N}_{m,k}\right),
\end{equation}
where $\mathbf{S}_{m,k}=\left[ \mathbf{s}_{m,k}\left[ 1 \right] ,\cdots ,\mathbf{s}_{m,k}\left[ N_{\text{s}} \right] \right] \in \mathbb{C} ^{N_{\text{MS}}\times N_{\text{s}}}$ and $\mathbf{N}_{m,k}$ denotes the received additive white Gaussian noise matrix at the $k$-th subcarrier in the $m$-th mini-slot. To simplify the design of systems, we assume that the pilot signals transmitted at each subcarrier are the same, i.e. $\mathbf{s}_{m,k}\left[ n \right]=\mathbf{s}\left[ n \right]$ and $\mathbf{S}_{m,k}=\mathbf{S}$. Then, the received signals by stacking all of the $Q_{\text{BS}}$ common RF combining vectors in the $m$-th mini-slot at the $k$-th subcarrier are given by
\begin{equation}
	\mathbf{Y}_{m,k}=\mathbf{W}^{\mathrm T}\left(\mathbf{H}_{m,k} \mathbf{S}+\mathbf{N}_{m,k}\right),
\end{equation}
where $\mathbf{W}=\left[ \mathbf{w}_1,\cdots ,\mathbf{w}_{Q_{\mathrm{BS}}} \right] \in \mathbb{C} ^{N_{\mathrm{BS}}\times Q_{\mathrm{BS}}}$. By stacking the signals of $K$ subcarriers at the $m$-th mini-slot, the received signals can be expressed as a three way tensor $\mathcal{Y}_m\in\mathbb{C} ^{Q_{\mathrm{BS}}\times N_{\mathrm{s}}\times K}$
\begin{equation}
	\mathcal{Y} _m=\left(\sum_{l=1}^L\left( \mathbf{a}_{N_{\mathrm{BS}}}\left( \theta _{l}\right) \circ \mathbf{S}^{\mathrm T}\mathbf{a}_{N_{\mathrm{MS}}}\left( \phi_{l} \right) \circ \tilde{\mathbf{\bf c}}_l\right)+\mathcal{N}_m\right)  \times _1{\bf W},
\end{equation}
where $\tilde{\mathbf{\bf c}}_l=e^{j2\pi f_{l}^{\text d}\left( m-1 \right) N_{\text{s}}T_{\text{s}}}{\bf c}_l$, and $\mathcal{N}_m\in\mathbb{C} ^{N_{\mathrm{BS}}\times N_{\text{s}}\times K}$ is formed by stacking all the noise matrices $\mathbf{N}_{m,k}$ from $K$ subcarriers in the $m$-th mini-slot. By stacking all the signals at $K$ subcarriers in $M$ mini-slots, a four way tensor $\mathcal{Y}\in\mathbb{C} ^{Q_{\text{BS}}\times N_{\text{s}}\times K \times M}$ can be formed, where its CP decomposition is expressed as 
\begin{equation}
	\mathcal{Y} =\left(\sum_{l=1}^L\left( \mathbf{a}_{N_{\mathrm{BS}}}\left( \theta _{l}\right) \circ \mathbf{S}^{\mathrm T}\mathbf{a}_{N_{\text{MS}}}\left( \phi_{l} \right) \circ {\mathbf{\bf c}}_l \circ {\mathbf{\bf d}}_l \right)+\mathcal{N}\right)  \times _1{\bf W}.
\end{equation}
Denoting $\mathcal{N}^{\bf W}=\mathcal{N}\times _1{\bf W}$, $\mathcal{Y}$ can be also expressed in terms of its PARAFAC decomposition
\begin{equation}\label{observeY}
	\mathcal{Y} =\mathcal{I}_{4,L}\times _1{\mathbf{A}}\times _2{\mathbf{B}}\times _3{\mathbf{C}}\times_4{\mathbf{D}}+\mathcal{N}^{\bf W},
\end{equation}
where the factor matrices $\bf A$ and $\bf B$ are given by
\begin{align}
	{\mathbf{A}}&\triangleq\mathbf{W}^{\mathrm T}{\bf A}_{N_\text{BS}},
	\\
	{\mathbf{B}}&\triangleq\mathbf{S}^{\mathrm T}{\bf A}_{N_\text{MS}},
\end{align}
and the factor matrices $\mathbf{C}=\left[ \mathbf{c}_1,\mathbf{c}_2,\cdots ,\mathbf{c}_L \right] $ and $\mathbf{D}=\left[ \mathbf{d}_1,\mathbf{d}_2,\cdots ,\mathbf{d}_L \right] $, where $\mathbf{c}_l$ and $\mathbf{d}_l$ are defined in (\ref{cl}) and (\ref{dl}), respectively.

\vspace{-0.5cm}
\section{Tensor Decomposition Based Channel Estimation Methods}\label{channelestimation}
In this section, we introduce the proposed tensor decomposition based channel estimation algorithms for time-varying mmWave MIMO-OFDM systems. First, we discuss the recovery of the four factor matrices from the noisy signals by solving the CP decomposition problem, and an ESPRIT-type decomposition based method is proposed. Then, the channel parameters are estimated based on the factor matrices. Finally, the uniqueness condition of the CP decomposition for the proposed model is discussed. 
\vspace{-0.5cm}

\subsection{ESPRIT-type Decomposition Based Channel Estimation}
\subsubsection{Factor Matrices Recovery}
The factor matrices can be recovered by solving the  CP decomposition problem
\begin{equation}\label{CPprob}
	\min_{\mathbf{A},\mathbf{B},\mathbf{C},\mathbf{D}} \left\| \mathcal{Y} -\sum_{l=1}^L{\mathbf{a}_l\circ \mathbf{b}_l\circ \mathbf{c}_l\circ \mathbf{d}_l} \right\|_{F}^{2}.
\end{equation}

Denote $\mathcal{X}= [\kern-0.15em[ \left( \mathbf{B}\odot \mathbf{A} \right) , \mathbf{C}, \mathbf{D}]\kern-0.15em] + \mathcal{N}^{\bf W}_{\text r}\in\mathbb{C}^{N_{\text s}Q_{\text{BS}}\times K \times M} $ as the reshaped version of four way tensor ${\mathcal{Y} }= [\kern-0.15em[ \mathbf{A} ,  \mathbf{B} , \mathbf{C}, \mathbf{D}]\kern-0.15em] $, where $\mathcal{N}^{\bf W}_{\text r}$ is the reshaped version of the noise tensor $\mathcal{N}$. In this section, we solve the CP decomposition problem by leveraging the structure of the Vandermonde matrix \cite{Vandermonde}. 

The mode-1 unfolding of $\mathcal{X}$ is given by $\mathbf{X}_1=\left( \mathbf{D}\odot \mathbf{C} \right) \left( \mathbf{B}\odot \mathbf{A} \right) ^{\mathrm T}\in \mathbb{C}^{MK\times Q_{\text{BS}}N_{\text s}} $. Denote an integer pair $(K_4,L_4)$ that satisfies $K_4+L_4 = M+1$, and define $\mathbf{J}_l\triangleq\left[ \begin{matrix}
	\mathbf{0}_{K_4\times \left( l-1 \right)}&		\mathbf{I}_{K_4}&		\mathbf{0}_{K_4\times \left( L_4-l \right)}\\
\end{matrix} \right] \in \mathbb{C} ^{K_4\times M}$. Then, the spatial smoothing of $\mathbf{X}_1$ is defined as \cite{Vandermonde}
\begin{align}\label{Xs1}
\mathbf{X}_s&=\left[ \left( \mathbf{J}_1\otimes \mathbf{I}_K \right) \mathbf{X}_1,		\cdots,		\left( \mathbf{J}_{L_4}\otimes \mathbf{I}_{K} \right) \mathbf{X}_1\right] \nonumber\\
&=\left[ \left( \mathbf{J}_1\otimes \mathbf{I}_K \right) \left( \mathbf{D}\odot \mathbf{C} \right) \left( \mathbf{B}\odot \mathbf{A} \right) ^{\mathrm{T}},\cdots,\right.\nonumber\\
&\qquad\qquad\qquad \left.\left( \mathbf{J}_{L_4}\otimes \mathbf{I}_K \right) \left( \mathbf{D}\odot \mathbf{C} \right) \left( \mathbf{B}\odot \mathbf{A} \right) ^{\mathrm{T}}
\right].
\end{align}
By using the property of Khatri-Rao product $(\mathbf{A}\odot \mathbf{B})(\mathbf{C}\otimes \mathbf{D})=(\mathbf{AC})\odot (\mathbf{BD})$ \cite{zxd}, Equation (\ref{Xs1}) can be rewritten as
\begin{align}\label{Xs}
&\mathbf{X}_s=\left[\left( \left( \mathbf{J}_1\mathbf{D} \right) \odot \mathbf{C} \right) \left( \mathbf{B}\odot \mathbf{A} \right) ^{\mathrm{T}},\cdots,\right.\nonumber\\
&\qquad\qquad\qquad\left.\left( \left( \mathbf{J}_{L_4}\mathbf{D} \right) \odot \mathbf{C} \right) \left( \mathbf{B}\odot \mathbf{A} \right) ^{\mathrm{T}}\right].
\end{align}
Note that the factor matrix $\bf D$ is a Vandermonde matrix, and thus we have 
\begin{equation}
	\mathbf{J}_l\mathbf{D}=\mathbf{D}^{K_4}\mathrm{D}_l\left( \mathbf{D} \right) ,
\end{equation}
where $\mathbf{D}^{K_4}$ denotes the first $K_4$ rows of $\mathbf{D}$. By leveraging the Vandermond structure of factor matrix $\bf D$, we have 
\begin{align}
	{\bf X}_s&=\left[
	\left( \mathbf{D}^{K_4}\mathrm{D}_1\left( \mathbf{D} \right) \odot \mathbf{C} \right) \left( \mathbf{B}\odot \mathbf{A} \right) ^{\mathrm{T}},\cdots,\right.\nonumber\\
	&\qquad\qquad\qquad\left.\left( \mathbf{D}^{K_4}\mathrm{D}_{L_4}\left( \mathbf{D} \right) \odot \mathbf{C} \right) \left( \mathbf{B}\odot \mathbf{A} \right) ^{\mathrm{T}}\right] \nonumber\\
	&=\left[	\left( \mathbf{D}^{K_4}\odot \mathbf{C} \right) \mathrm{D}_1\left( \mathbf{D} \right) \left( \mathbf{B}\odot \mathbf{A} \right) ^{\mathrm{T}},\cdots,\right.\nonumber\\
	&\qquad\qquad\qquad\left.\left( \mathbf{D}^{K_4}\odot \mathbf{C} \right) \mathrm{D}_{L_4}\left( \mathbf{D} \right) \left( \mathbf{B}\odot \mathbf{A} \right) ^{\mathrm{T}} \right] 
	\nonumber\\
	&=\left( \mathbf{D}^{K_4}\odot \mathbf{C} \right) \left( \mathbf{D}^{L_4}\odot \left( \mathbf{B}\odot \mathbf{A} \right) \right) ^{\mathrm{T}}.
\end{align}
With the known structure of ${\bf X}_s$, we can recover the factor matrices based on the subspace decomposition method and the estimation theory. Based on the assumption that the angles and the Doppler shifts of different paths are unequal, the factor matrices are full column rank. Denote the singular value decomposition (SVD) of ${\bf X}_s$ as ${\bf X}_s = \mathbf{U\Sigma V}^{\mathrm H}$, where the left singular matrix ${\bf U}$ is the column space of ${\bf Y}_s$ that contains the basis of $\left( \mathbf{D}^{K_4}\odot \mathbf{C} \right)$. Assuming that the number of paths $L$ is known, there exists a full rank matrix ${\bf M}\in\mathbb{C}^{L\times L}$ that
\begin{equation}
	\mathbf{U}_{:,1:L}\mathbf{M}=\left( {\mathbf{D}}^{K_4}\odot {\mathbf{C}} \right) .
\end{equation}
Define
\begin{align}
\mathbf{U}_1&=\left[ \mathbf{U} \right] _{1:\left( K_4-1 \right) K,1:L\,\,}\in \mathbb{C} ^{\left( K_4-1 \right) K\times L},\\
\mathbf{U}_2&=\left[ \mathbf{U} \right] _{K+1:K_4K,1:L}\in \mathbb{C} ^{\left( K_4-1 \right) K\times L}.
\end{align} 
Then, we have 
\begin{align}
	\mathbf{U}_1\mathbf{M}&=\left( \underline{\mathbf{D}}^{K_4}
	\odot {\mathbf{C}} \right) \label{U1}, \\
	\mathbf{U}_2\mathbf{M}&=\left( \overline{{\mathbf{D}}}^{K_4}\odot {\mathbf{C}} \right) =\mathbf{U}_1\mathbf{MZ},\label{U2}
\end{align}
where $\underline{\mathbf{D}}^{K_4}$ and $\overline{\mathbf{D}}^{K_4}$ are the ${\mathbf{D}}^{K_4}$ that deletes the last row and the first row, respectively, and ${\bf Z}$ is a diagonal matrix, the diagonal elements are the generator of factor ${\mathbf D}$, ${\bf Z}\triangleq {\mathrm{D}\left(\left[e^{j2\pi f^{d}_{1}N_{\mathrm s}T_{\mathrm s}},\cdots,e^{j2\pi f^{d}_{L}N_{\mathrm s}T_{\mathrm s}}\right]^{\mathrm T}\right)}$. Since the property of the Vandermonde matrix $\overline{\mathbf{D}}^{K_4}=\underline{\mathbf{D}}^{K_4}{\bf Z}$, by combining (\ref{U1}) with (\ref{U2}), we have
\begin{equation}
	\mathbf{U}_{1}^{\dagger}\mathbf{U}_2=\mathbf{MZM}^{-1}=\mathbf{P\Lambda P}^{-1},
\end{equation}
where $\mathrm{d}\left( \mathbf{\Lambda } \right) =\mathrm{d}\left( \mathbf{Z} \right) \mathbf{\Pi }$ and $\mathbf{P\Lambda P}^{-1}$ is the eigenvalue decomposition (EVD) of $\mathbf{U}_{1}^{\dagger}\mathbf{U}_2$, where $\mathbf{P}=\mathbf{M\Delta \Pi }$ is similar to ${\bf M}$, and $\mathbf{\Pi }$ is a permutation matrix, $\mathbf{\Delta}$ is a diagonal scaling ambiguity matrix. From the EVD of $\mathbf{U}_{1}^{\dagger}\mathbf{U}_2$, we can estimate the generator of ${\bf D}$, and thus recover the ambiguity version of factor $\bf D$ as
\begin{equation}
	\widehat{\mathbf{D}}={\mathbf{D}}\mathbf{\Pi }.
\end{equation}
Next, we recover factor matrix $\bf C$ according to the recovered $\bf D$ with permutation ambiguity and the eigenvectors $\bf P$. By using the similarity $\mathbf{P}=\mathbf{M\Delta \Pi }$, we have 
\begin{align}
\mathbf{UP}&=\mathbf{UM\Delta \Pi }\nonumber\\
&=\left( {\mathbf{D}}^{K_4}\odot {\mathbf{C}} \right) \mathbf{\Delta \Pi }\nonumber\\
&=\left( {\mathbf{D}}^{K_4}\mathbf{\Pi }\odot {\mathbf{C}}\mathbf{\Pi \Delta } \right)\nonumber\\
&=\left( \widehat{\mathbf{D}}^{K_4}\odot \widehat{\mathbf{C}} \right) ,
\end{align}
where $\widehat{\mathbf{D}}^{K_4}$ and $\widehat{\mathbf{C}}$ are the ${\mathbf{D}}^{K_1}$ with permutation ambiguity and ${\mathbf{C}}$ with permutation and scaling ambiguity, respectively. Thus, we have
\begin{equation}
	\mathbf{UP}_{:,l}=\left( \widehat{\mathbf{d}}_{l}^{K_4}\otimes \widehat{\mathbf{c}}_l \right) ,
\end{equation}
and $\widehat{\mathbf{c}}_l$ can be estimated by using the LS estimation
\begin{equation}\label{hatC}
\widehat{\mathbf{c}}_l=\frac{\left( \widehat{\mathbf{d}}_{l}^{K_4} \right) ^{\mathrm H}\otimes \mathbf{I}_K}{\left( \widehat{\mathbf{d}}_{l}^{K_4} \right) ^{\mathrm H}\widehat{\mathbf{d}}_{l}^{K_4}}\mathbf{Up}_l.
\end{equation}
Hence, we obtain the factor $\widehat{\mathbf{D}}$ and $\widehat{\mathbf{C}}$ with ambiguity. Then, we recover the factor $\bf A$ and $\bf B$ from the known $\widehat{\mathbf{D}}$ and $\widehat{\mathbf{C}}$ and the SVD of ${\bf X}_{\text s}$. The procedure of the recovery of ${\mathbf{B}}\odot{\mathbf{A}}$ is similar to that of $\bf C$. By exploiting the right singular vectors $\bf V$, we have
\begin{equation}\label{V1}
	\mathbf{V}^{\ast}\mathbf{\Sigma P}^{-{\mathrm T}}=\left( {\mathbf{D}}^{L_4}\odot {\mathbf{B}}\odot{\mathbf{A}} \right). 
\end{equation}
Substituting $\mathbf{P}=\mathbf{M\Delta \Pi }$ into (\ref{V1}), we have 
\begin{align}
	\mathbf{V}^{\ast}\mathbf{\Sigma P}^{-{\mathrm T}}&=\mathbf{V}^{\ast}\mathbf{\Sigma M}^{-{\mathrm T}}\mathbf{\Delta }^{-1}\mathbf{\Pi }\nonumber\\
	&=\left( {\mathbf{D}}^{L_4}\odot {\mathbf{B}}\odot {\mathbf{A}} \right) \mathbf{\Delta }^{-1}\mathbf{\Pi }
	\nonumber\\
	&=\left({\mathbf{D}}^{L_4}\mathbf{\Pi }\odot \left( {\mathbf{B}}\odot {\mathbf{A}} \right) \mathbf{\Delta }^{-1}\mathbf{\Pi } \right) \nonumber\\
	&=\left(\widehat{\mathbf{D}}^{L_4}\odot \left( \widehat{\mathbf{B}}\odot \widehat{\mathbf{A}} \right) \right) ,
\end{align}
where $\widehat{\mathbf{D}}^{L_4}$ is ${\mathbf{D}}^{L_4}$ with permutation ambiguity, and $\widehat{\mathbf{A}}$ and $\widehat{\mathbf{B}}$ are the factor matrices $\bf A$ and $\bf B$ with permutation and scaling ambiguity. Letting $\widehat{\mathbf{E}}=\left( \widehat{\mathbf{B}}\odot \widehat{\mathbf{A}} \right) $ and $\mathbf{N}=\mathbf{P}^{-{\mathrm T}}$, we have 
\begin{equation}\label{hatE}
\widehat{\mathbf{e}}_l=\frac{\left( \widehat{\mathbf{d}}_{l}^{L_4} \right) ^{\mathrm{H}}\otimes \mathbf{I}_{Q_{\text{BS}}N_{\text{s}}}}{\left( \widehat{\mathbf{d}}_{l}^{L_4} \right) ^{\mathrm{H}}\widehat{\mathbf{d}}_{l}^{L_4}}\mathbf{V}^{\ast}\mathbf{\Sigma n}_l.
\end{equation}
Hence, we obtain the ambiguous version of $\bf E$. The recoveries of $\bf A$ and $\bf B$ are as follows. Let $\widehat{\mathbf{E}}_l=\mathrm{unvec}_{Q_{\text{BS}}\times N_{\text{s}}}\left( \widehat{\mathbf{e}}_l \right) $, then $\bf A$ and $\bf B$ can be estimated by solving the following problem
\begin{equation}\label{AB}
\mathrm{arg} \underset{\widehat{\mathbf{a}}_l,\widehat{\mathbf{b}}_l}{\min}\left\| \widehat{\mathbf{E}}_l-\mathbf{a}_l\mathbf{b}_{l}^{\mathrm{T}} \right\| _{F}^{2}.
\end{equation}
The problem can be solved from SVD of $\widehat{\mathbf{E}}_l$. Let $\hat{\mathbf{a}}_l=\lambda _{l,1}\hat{\mathbf{u}}_{l,1}$ and $\hat{\mathbf{b}}_l=\hat{\mathbf{v}}_{l,1}^{\ast}$, where $\lambda_l$, $ \hat{\mathbf{u}}_{l,1}$ and $\hat{\mathbf{v}}_{l,1}$ are the maximum singular value of $\widehat{\mathbf{E}}_l$, the corresponding left singular vector and the corresponding right singular vector, respectively. Then, we obtain the ambiguity versions of $\bf A$ and $\bf B$.

\subsubsection{Parameters Estimation}\label{paraesti}

Based on the recovered factor matrices, we can estimate the channel parameters. Due to the ambiguity of the tensor decomposition, the four factor matrices are permuted by the same permutation matrix $\bf \Pi$ and scaled by different scale matrices, i.e.,
\begin{subequations}
	\begin{align}
		\widehat{\mathbf{A}}&=\mathbf{A\Delta }_1\mathbf{\Pi }+\mathbf{E}_1
		\\
		\widehat{\mathbf{B}}&=\mathbf{B\Delta }_2\mathbf{\Pi }+\mathbf{E}_2
		\\
		\widehat{\mathbf{C}}&=\mathbf{C\Delta }_3\mathbf{\Pi }+\mathbf{E}_3
		\\
		\widehat{\mathbf{D}}&=\mathbf{D\Delta }_4\mathbf{\Pi }+\mathbf{E}_4,
	\end{align}
\end{subequations}
where $\mathbf{\Delta }_i$ and the $\mathbf{E}_i$ are the scaling matrix and the estimation error of the $i$-th factor matrix, respectively. However, the ambiguity does not affect the estimation of angles and delays since the scaling and permutation do not affect the direction of the columns of the factor matrices in the signal space, and thus the estimations of angles and the delays are not affected. First, the Doppler shifts can be estimated by
\begin{equation}\label{estFd}
	\widehat{f}_{l}^{\text d}=\frac{1}{2\pi N_{\text s}T_{\text s}}\angle \left( \lambda_l \right) ,
\end{equation}
where $\lambda_l$ is the $l$-th element of ${\text d}({\bf \Lambda})$ and $\angle$ denotes the phase calculation operator.

Then, we estimate the AoAs and the AoDs. Since the hybrid precoding architecture is applied instead of fully-digital architecture, the precoding matrix ${\bf F}_{\text {RF}}$ and the combining matrix ${\bf W}$ are row rank deficient and the LS estimate is not applicable. Instead, a correlation based angle estimation scheme can be applied
\begin{align}
	\widehat{\theta} _l&=\mathrm{arg}\underset{\theta}{\max}\frac{\left| \widehat{\mathbf{a}}_{l}^{\mathrm{H}}\mathbf{W}^{\mathrm{T}}\mathbf{a}_{\text{BS}}\left( \theta \right) \right|^2}{\left\| \widehat{\mathbf{a}}_l \right\| ^{2}\left\| \mathbf{W}^{\mathrm{T}}\mathbf{a}_{\text{BS}}\left( \theta \right) \right\| ^{2}},\label{estAOA}
	\\
	\widehat{\phi} _l&=\mathrm{arg}\underset{\phi}{\max}\frac{\left| \widehat{\mathbf{b}}_{l}^{ \mathrm{H}}\mathbf{S}^{\mathrm{T}}\mathbf{a}_{\text{MS}}\left( \phi \right) \right|^2}{\left\| \widehat{\mathbf{b}}_l \right\| ^{2}\left\| \mathbf{S}^{\mathrm{T}}\mathbf{a}_{\text{MS}}\left( \phi \right) \right\| ^{2}},\label{estAOD}
\end{align}
where $\widehat{\mathbf{a}}_{l}$ and $\widehat{\mathbf{b}}_{l}$ are the $l$-th column of the estimated factor matrices $\widehat{\mathbf{A}}$ and $\widehat{\mathbf{B}}$, respectively. 

The delays $\{\tau_l\}$ can be estimated as follows. Since the factor $\bf C$ can be rewritten as a Vandermonde matrix multiplying a scaling matrix
\begin{equation}
\mathbf{C}=\overline{\mathbf{C}}\mathrm{D}\left( \left[\alpha _1e^{j{2\pi}f_{1}^{\text{d}}\tau _1},\cdots ,\alpha _Le^{j{2\pi}f_{L}^{\text{d}}\tau _L}\right]^{\mathrm T} \right) ,
\end{equation}
where 
$$
\overline{\mathbf{C}}_{:,l}=\left[ e^{-j\frac{2\pi}{N}\left( f_{\text s}\tau _l \right)},\cdots ,e^{-j\frac{2\pi}{N}K\left( f_{\text s}\tau _l \right)} \right] ^{\mathrm{T}}.
$$
Thus we have 
\begin{align}
	\overline{{\mathbf{C}}}&=\underline{{\mathbf{C}}}\mathbf{Z}_c,\\
	\overline{\widehat{\mathbf{C}}}&=\underline{\mathbf{C}}\mathbf{Z}_{\mathrm{c}}\mathbf{\Delta }_3\mathbf{\Pi },
\end{align}
where ${\bf Z}_{\mathrm c}$ is a diagonal matrix and the principal diagonal elements are the generator of the Vandermonde matrix $\overline{\bf C}$. Hence, the delays $\{\tau_l\}$ can be estimated as 
\begin{equation}
	\widehat{\tau}_l=\frac{N}{-2\pi f_{\text s}}\angle \left(\left( \underline{\widehat{\mathbf{c}}}_l \right) ^{\dagger}\overline{\widehat{\mathbf{c}}}_l\right).\label{estdelay}
\end{equation}

Finally, we estimate the scaling factor and the channel gain $\{\alpha_l\}$. Due to the property of the ambiguity of tensor decomposition, we first estimate the scaling ambiguity 
\begin{subequations}
	\begin{align}
		\left( \mathbf{\Delta }_1 \right) _{l,l}&=\left( \mathbf{W}^{\mathrm{T}}\mathbf{a}_{\text{BS}}\left( \hat{\theta}_l \right) \right) ^{\dagger}\widehat{\mathbf{a}}_l,\label{delta1} \\
		\left( \mathbf{\Delta }_2 \right) _{l,l}&=\left( \mathbf{S}^{\mathrm{T}}\mathbf{a}_{\text{MS}}\left( \hat{\phi}_l \right) \right) ^{\dagger}\hat{\mathbf{b}}_{l},\label{delta2}\\
		\left( \mathbf{\Delta }_4 \right) _{l,l}&=\mathbf{d}^{\dagger}\left( \widehat{f}_{l}^{\text{d}} \right) \widehat{\mathbf{d}}_{l},\label{delta4}\\
		\mathbf{\Delta }_3&=\mathbf{\Delta }_{1}^{-1}\mathbf{\Delta }_{2}^{-1}\mathbf{\Delta }_{4}^{-1},\label{delta3}
	\end{align}
\end{subequations}
where $\mathbf{d}\left( f_{l}^{\text{d}} \right) \triangleq\left[ 
	1,\cdots,e^{j2\pi f_{l}^{\text{d}}\left( M-1 \right) N_{\text{s}}T_{\text{s}}
	}\right]^{\mathrm T}$. Substituting (\ref{delta3}) into (\ref{cl}), we have the LS estimator of $\alpha_l$ as
\begin{equation}
\hat{\alpha}_l=\left( \mathbf{\Delta }_{3}^{-1} \right) _{l,l}\bar{\mathbf{c}}\left( \tau _l \right) ^{\dagger}{\mathbf{c}}_le^{-j2\pi \left( f_{l}^{\text{d}}\widehat{\tau}_l \right)},\label{estalpha}
\end{equation}
where $\mathbf{c}\left( \tau _l \right) =\left[ e^{-j\frac{2\pi}{N}\left( f_{\text s}\tau _l \right)},\cdots,e^{-j\frac{2\pi}{N}K\left( f_{\text s}\tau _l \right)}\right] ^{\mathrm T}$, and $\bar{\mathbf{c}}\left( \tau _l \right)$ is obtained by removing the first element of  $\mathbf{c}\left( \tau _l \right)$. The proposed algorithm is summarized in Algorithm \ref{alg:1}.

\begin{algorithm}[h]
	\caption{ESPRIT-type Decomposition based Time Varying Channel Estimation Algorithm}
	\label{alg:1}
	\begin{algorithmic}[1]
		
		\REQUIRE $\mathcal{Y}$
		\ENSURE $\{\widehat{\theta}_l\}$, $\{\widehat{\phi}_l\}$, $\{\widehat{\tau}_l\}$, $\{\widehat{\alpha}_l\}$, $\{\widehat{f}^{\text d}_l\}$
		\STATE Choose integer pair $(K_4,L_4)$ that satisfies $K_4+L_4=M+1$. Compute the reshaped version $\mathcal{X}= [\kern-0.15em[ \left( \mathbf{B}\odot \mathbf{A} \right) , \mathbf{C}, \mathbf{D}]\kern-0.15em] + \mathcal{N}^{\bf W}_{\text r}\in\mathbb{C}^{Q_{\text{BS}}N_{\text s}\times K \times M} $.
		\STATE Compute the spatial smooth ${\bf X}_s$ by (\ref{Xs}). Compute the SVD ${\bf X}_s = \mathbf{U\Sigma V}^{\mathrm H}$. \\
		\STATE Compute the EVD $\mathbf{U}_{1}^{\dagger}\mathbf{U}_2=\mathbf{P\Lambda P}^{-1}$, where $\mathbf{U}_{1}$ and $\mathbf{U}_2$ are defined in (\ref{U1}) and (\ref{U2}), respectively.
		\STATE Estimate the generators $\widehat{z}_l={\bf Z}_{l,l}/{|{\bf Z}_{l,l}|}$ and reconstruct $\widehat{\bf D}$ as $\widehat{\bf d}_l=[\widehat{z}_l,\widehat{z}^2_l,\cdots,\widehat{z}^M_l]^{\mathrm T} $.
		\STATE Compute ${\bf UP}$ and reconstruct $\widehat{\bf C}$ by (\ref{hatC}).
		\STATE Compute ${\bf P}^{-{\mathrm T}}$ and reconstruct $\widehat{\bf B}$ and $\widehat{\bf A}$ by (\ref{hatE}) and (\ref{AB}), respectively. 
		\STATE Estimate the Doppler shift $\{f^{\text d}_l\}$, AoAs $\{{\theta}_l\}$, AoDs $\{{\phi}_l\}$ and delays $\{{\tau}_l\}$ by (\ref{estFd}), (\ref{estAOA}), (\ref{estAOD}) and (\ref{estdelay}), respectively. 
		\STATE Estimate the scaling ambiguity $\{{\bf \Delta}_i\}$ by (\ref{delta1})-(\ref{delta3}), respectively and estimate the channel gain $\{ \alpha_l \}$  by (\ref{estalpha}).   
		\STATE {\bf return} channel parameters $\{\widehat{f}^{\text d}_l, \widehat{\theta}_l, \widehat{\phi}_l, \widehat{\tau}_l,  \widehat{\alpha}_l \}$ and channel matrices $\{\widehat{\bf H}_{m,k}\}$.                       
	\end{algorithmic}
\end{algorithm}
\vspace{-0.5cm}

\subsection{Uniqueness}
In this subsection, we discuss the uniqueness of the CP decomposition. 
For a general $N$-th-order tensor, a unique condition is given in \cite{uniqueness}:

{\textbf{Lemma 1}} (Uniqueness of multilinear decomposition \cite{uniqueness}): Letting $\mathcal{X} =[\kern-0.15em[ \mathbf{A}^{\left( 1 \right)},\mathbf{A}^{\left( 2 \right)},\cdots ,\mathbf{A}^{\left( N \right)} ]\kern-0.15em]
$ be a CP solution which decomposes an $N$th-order tensor $\mathcal{X} \in \mathbb{C} ^{I_1\times I_2\times \cdots \times I_N}$ into $R$ rank-one arrays, where $\mathbf{A}^{\left( 1 \right)}\in\mathbb{C}^{I_1\times R}$, $\mathbf{A}^{\left( 2 \right)}\in\mathbb{C}^{I_2\times R}$, $\cdots$, $\mathbf{A}^{\left( N \right)}\in\mathbb{C}^{I_N\times R}$, the solution is unique if
\vspace{-0.1cm}
{\setlength\abovedisplayskip{3pt}
	\setlength\belowdisplayskip{3pt}
	\begin{equation}\label{aproexpression}
		\sum_{n=1}^Nk\left({{\mathbf{A}^{\left( n \right)}}}\right)\geqslant 2R+N-1.
\end{equation}}

Though Lemma 1 gives a general uniqueness condition for arbitrary $N$th-order tensor, by leveraging the property of spatial smoothing and Vandermonde structure, a more relaxed uniqueness condition can be derived. The Corollary III.4 in \cite{Vandermonde} given a uniqueness condition of the third-order CP decomposition under the spatial smoothing and ESPRIT algorithm, where two of three factor matrices are assumed to be Vandermonde matrices. In this section, we present the uniqueness condition of fourth-order CP decomposition as follows.

{\textbf{Remark 1}}: Let $\mathcal{X} \in \mathbb{C} ^{I_1\times I_2\times I_3\times I_4}$ be a fourth-order tensor with factor matrices $\mathbf{A}^{\left( n \right)}\in \mathbb{C} ^{I_n\times R}, n\in \left\{ 1,2,3,4 \right\} $, where $\mathbf{A}^{\left( 1 \right)}$ and $\mathbf{A}^{\left( 2 \right)}$ are Vandermonde matrices with distinct generators $\left\{ z_{1,r} \right\} _{r=1}^{R}$ and $\left\{ z_{2,r} \right\} _{r=1}^{R}$, respectively. Consider the the reshaped version of $\mathcal{X} $ as $\mathcal{X} _{\text{r}}=[\kern-0.15em[ \left( \mathbf{A}^{\left( 2 \right)}\odot \mathbf{A}^{\left( 1 \right)} \right) ,\mathbf{A}^{\left( 3 \right)},\mathbf{A}^{\left( 4 \right)} ]\kern-0.15em] \in \mathbb{C} ^{I_2I_1\times I_3\times I_4}$, and matrix representation $\left( \mathbf{A}^{\left( K_1,2 \right)}\odot \mathbf{A}^{\left( 1 \right)} \right) \left( \mathbf{A}^{\left( L_1,2 \right)}\odot \mathbf{A}^{\left( 4 \right)}\odot \mathbf{A}^{\left( 3 \right)} \right) ^{\mathrm{T}}$ with $K_1+L_1=I_1+1$. If
\begin{equation}\label{uniqueness1}
\left\{ \begin{aligned}
	&r\left( {\mathbf{A}}^{\left( K_1,2 \right)}\odot \mathbf{A}^{\left( 1 \right)} \right) =R\\
	&r\left( {\mathbf{A}}^{\left( L_1,2 \right)}\odot \mathbf{A}^{\left( 4 \right)}\odot \mathbf{A}^{\left( 3 \right)} \right) =R,\\
\end{aligned} \right.  
\end{equation}
then the rank of $\mathcal{X} $ is $R$ and the Vandermonde constrained CPD of $\mathcal{X} $ is unique. Condition (\ref{uniqueness1}) is generally satisfied if
\begin{equation}\label{uniqueness1}
	\left\{ \begin{aligned}
		&\left( K_1-1 \right) I_2\geqslant R\\
		&	k\left( \mathbf{A}^{\left( 3 \right)} \right) +k\left( \mathbf{A}^{\left( 4 \right)} \right) \geqslant \left\lceil \frac{R}{L_1} \right\rceil +1.,\\
	\end{aligned} \right.  
\end{equation}

{\textbf{Proof}}: Let  $\mathbf{X}=\mathbf{U\Sigma V}^{\mathrm{H}}$ denote the SVD of $\mathbf{X}$. Under the conditions in Proposition 1, there exists a nonsingular matrix $\mathbf{P}\in \mathbb{C} ^{R\times R}$ satisfying 
\begin{equation}
	\mathbf{UP}=\mathbf{A}^{\left( K_1,1 \right)}\odot \mathbf{A}^{\left( 2 \right)}.
\end{equation}
By using Algorithm 1, $\mathbf{A}^{\left( K_1,1 \right)}$ and its generators $z_{1,r}$ can be firstly found. Then, $\mathbf{A}^{\left( 2 \right)}$ can be computed as
\begin{equation}
	\mathbf{a}_{r}^{\left( 2 \right)}=\left( \mathbf{a}_{r}^{\left( K_1,1 \right) \mathrm{H}}\otimes \mathbf{I}_{I_2} \right) \mathbf{Up}_r.
\end{equation} 
Considering the fact that $\mathbf{A}^{\left( 2 \right)}$ is also Vandermonde matrix, its generators $z_{2,r}$ can be computed as $z_{2,r}=\left( \underline{\mathbf{a}}_{r}^{\left( 2 \right)} \right) ^{\dagger}\overline{\mathbf{a}}_{r}^{\left( 2 \right)}$. The next steps are to recover the factors $\mathbf{A}^{\left( 3 \right)}$ and $\mathbf{A}^{\left( 4 \right)}$. Note that the factors $\mathbf{A}^{\left( 3 \right)}$ and $\mathbf{A}^{\left( 4 \right)}$ are coupled as $\mathbf{E}=\mathbf{A}^{\left( 3 \right)}\odot \mathbf{A}^{\left( 4 \right)}$. When the CP decomposition is unique, the condition $r\left( {\mathbf{A}}^{\left( L_1,2 \right)}\odot \mathbf{E} \right) \geqslant R$ holds. Since the Vandermonde structure of ${\mathbf{A}}^{\left( L_1,2 \right)}$, we have $k\left( \mathbf{A}^{\left( L_1,2 \right)}\odot \mathbf{E} \right) =\min \left( L_1k\left( \mathbf{E} \right) ,R \right) $. To ensure the uniqueness of the CP decomposition under this coupling relationship, the following lemma is introduced:

{\textbf{Lemma 2}} (k-rank of Khatri-Rao Product\cite{KRrank}): Given $\mathbf{A}\in \mathbb{C} ^{M\times R}$ and $\mathbf{B}\in \mathbb{C} ^{N\times R}$ satisfying $k\left( \mathbf{A} \right) \geqslant 1$ and $k\left( \mathbf{B} \right) \geqslant 1$, then we have $k\left( \mathbf{B}\odot \mathbf{A} \right) \geqslant \min \left( k\left( \mathbf{A} \right) +k\left( \mathbf{B} \right) -1,R \right) .$

Using Lemma 2, the inequality $r\left( {\mathbf{A}}^{\left( L_1,2 \right)}\odot \mathbf{E} \right) \geqslant R$ holds when
\begin{equation}
	k\left( \mathbf{A}^{\left( 3 \right)} \right) +k\left( \mathbf{A}^{\left( 4 \right)} \right) \geqslant \left\lceil \frac{R}{L_1} \right\rceil +1.
\end{equation} 
Based on the above discussions, we can derive the uniqueness condition (\ref{uniqueness1}). 
Compared to Lemma 1, Remark 1 gives a more relaxed uniqueness condition and does not prevent $k\left( \mathbf{A}^{\left( n \right)} \right) =1, n=1,2,3,4$.\hfill $\blacksquare$

\newcounter{TempEqCnt} 
\setcounter{TempEqCnt}{\value{equation}} 
\setcounter{equation}{62} 
\begin{figure*}[hb] 
	\hrulefill  
	\begin{align} \label{loglkh}
		L\left( \mathbf{p} \right) &=-Q_{\text{BS}}N_{\text s}KM\ln \left( \pi \sigma _{\text{n}}^{2} \right) -\ln  \det \left(\mathbf{W}^{\mathrm T}\mathbf{W}\otimes \mathbf{I}_{N_{\text s}KM} \right) -\left( \mathbf{y}_1-\mathbf{z}_1 \right) ^{\mathrm H}\left( \mathbf{C}_{\mathcal{N}} \right) ^{-1}\left( \mathbf{y}_1-\mathbf{z}_1 \right).	 
	\end{align}
\end{figure*}

\section{CRB Derivation}\label{CRB}
In this section, the CRB for estimating the channel parameters $\left\{ \theta _l,\phi _l,\tau _l,\alpha _l,f_{l}^{\text{d}} \right\} $ in Problem (\ref{CPprob}) is derived, which is the lower bound on the variance of any unbiased estimators \cite{kfs}. The observation tensor ${\mathcal{Y}}$ in (\ref{observeY}) can be rewritten as

\newcounter{TempEqCnt1} 
\setcounter{TempEqCnt1}{\value{equation}} 
\setcounter{equation}{60} 
\begin{equation}
	\mathcal{Y} =\underbrace{[\kern-0.15em[ \mathbf{A}, \mathbf{B}, \mathbf{C}, \mathbf{D} ]\kern-0.15em]}_{\triangleq\mathcal{Z}}+\mathcal{N} \times _1\mathbf{W}.
\end{equation}
Denote $\mathcal{N}^{\bf W}=\mathcal{N} \times _1\mathbf{W}$, and the corresponding mode-1 unfolding is $\mathbf{N}_{\left( 1 \right)}\mathbf{W}$, where ${\bf N}_{(1)}$ is the mode-1 unfolding of ${\mathcal{N}}$. Then, the vectorization of the mode-1 unfolding of $\mathcal{N} \times _1\mathbf{W}$ is given by
\begin{equation}
	\mathrm{vec}\left( \mathbf{N}_{\left( 1 \right)}\mathbf{W} \right) =\left( \mathbf{W}^{\mathrm T}\otimes \mathbf{I}_{N_{\text s}KM} \right) \mathrm{vec}\left( \mathbf{N}_{\left( 1 \right)} \right). 
\end{equation}
The covariance matrix of the noise vector $\mathrm{vec}\left( \mathbf{N}_{\left( 1 \right)}\mathbf{W} \right)$ is defined as  $\mathbf{C}_{\mathcal{N}}\triangleq\mathbb{E} \left\{ \mathrm{vec}\left( \mathbf{N}_{\left( 1 \right)}\mathbf{W} \right) \mathrm{vec}\left( \mathbf{N}_{\left( 1 \right)}\mathbf{W} \right) ^{\mathrm{H}} \right\} 
$. Based on the assumption that each element in $\mathcal{N}$ follows the i.i.d circular and symmetric complex Gaussian distribution $\mathcal{CN}(0,\sigma^2_{\text n})$, we have $
\mathbf{C}_{\mathcal{N}}=\sigma^2_{\text n}\left( \mathbf{W}^{\mathrm T}\mathbf{W}^{\ast} \right) \otimes \mathbf{I}_{N_{\text s}KM}
$. Then, denote $\mathbf{y}_1=\mathrm{vec}\left( \mathbf{Y}_{\left( 1 \right)} \right) $ and $\mathbf{z}_1=\mathrm{vec}\left( \mathbf{Z}_{\left( 1 \right)} \right) =\sum_{l=1}^L{\mathbf{a}_l\otimes \mathbf{d}_l\otimes \mathbf{c}_l\otimes \mathbf{b}_l}$ as the vectorized version of the mode-1 unfolding of $\mathcal{Y}$ and that of $\mathcal{Z}$, respectively. Define  the log-likelihood function $L({\bf p})$ in (\ref{loglkh}), which is at the bottom of this page, and $\mathbf{p}\triangleq\left[ \theta _1,\cdots ,\theta _L,\cdots ,f_{1}^{\text{d}},\cdots ,f_{L}^{\text{d}} \right] ^{\mathrm{T}}$. Then, the partial derivative of $L\left( \mathbf{p} \right)$ w.r.t. ${\bf p}$ can be derived. The Fisher information matrix (FIM) can then be computed as ${\bf I}({\bf p})=\mathbb{E}\left\{ \left( \frac{\partial L\left( \mathbf{p} \right)}{\partial \mathbf{p}} \right) ^H\left( \frac{\partial L\left( \mathbf{p} \right)}{\partial \mathbf{p}} \right) \right\} $, and the corresponding CRB can be obtained by computing the inverse of ${\bf I}({\bf p})$
\newcounter{TempEqCnt2} 
\setcounter{TempEqCnt2}{\value{equation}} 
\setcounter{equation}{63} 
\begin{equation}
	{\mathrm{CRB}}({\bf p})={\bf I}^{-1}({\bf p}).
\end{equation}
Further details for the derivations of the vectorization of ${\bf Z}_1$, the partial derivatives to $L({\bf p})$ and the derivations to ${\bf I}({\bf p})$ w.r.t. the corresponding channel parameters can be seen in the appendix.

\section{Complexity Analysis}\label{complexity}
In this section, the computational complexities of the proposed method is analyzed.
The complexity of the computation of ${\bf X}_s$ is on the order of $\mathcal{O}(K_4K^2MQ_{\text{BS}}N_{\text{s}}L_4)$. The SVD of ${\bf X}_s$ takes $\mathcal{O}(K_4K Q^2_{\text{BS}}N_{\text{s}}^2L^2_1 )$ flops. The multiplication $\mathbf{U}_{1}^{\dagger}\mathbf{U}_2$ has the complexity of $\mathcal{O}(L^2\left( K_4-1 \right) K)$. The complexity of the reconstruction of $\widehat{\bf C}$ is $\mathcal{O}(K^2K_4L)$. The reconstructions of $\widehat{\bf B}$ and $\widehat{\bf A}$ have the total complexity of $\mathcal{O}(K_4KL^2+Q_{\text{BS}}^{2}N_{\text{s}}L)$. The total complexity of the estimation of AOAs and AODs is $\mathcal{O}(Q_{\text{BS}}N_{\text{BS}}LG+N_{\text s}N_{\text{MS}}LG)$, where $G$ is associated with the precision of the angle searching grid. The estimation of $\{\tau_l\}$ takes $\mathcal{O}\left( K-1 \right) L)$ flops. Finally, the estimation of the scaling ambiguity and the channel gain $\{\alpha_l\}$ has the complexity of $\mathcal{O}(Q_{\text{BS}}N_{\text{BS}}L+N_{\text{s}}N_{\text{MS}}L+M+L+K)$. The total complexity is on the order of $\mathcal{O} ( K_4K^2MQ_{\text{BS}}N_{\text{s}}L_4+K_4K Q^2_{\text{BS}}N_{\text s}^2L^2_4  +Q_{\text{BS}}N_{\text{BS}}LG+N_{\text s}N_{\text{MS}}LG) $.

\section{Simulation Results}\label{simulation}

In this section, simulation results are presented to show the performance of the proposed channel estimation method and the comparison with benchmarks. The scales of the ULAs employed at the BS and the MS are $N_{\text{BS}}=128$ and $N_{\text{MS}}=64$, respectively. The number of RF chains at the BS and the MS are ${Q}_{\text{BS}}=16$ and ${Q}_{\text{BS}}=8$. The set of angular and delay parameters are generated from uniform distribution. The number of propagation paths is $L=3$. The complex channel gain $\alpha_l$ is generated from the complex Gaussian distribution of $\mathcal{C} \mathcal{N} \left( 0,\sigma _{\alpha}^{2}=( \frac{c}{4\pi Df_{\text{c}}} ) ^2 \right) $, where $D$ denotes the distance between the BS and the MS, and $f_{\text{c}}$ denotes the carrier frequency. In our considered scenario, we set $f_{\text{c}}=30$ GHz. The speed of the MS is set to $v_c=30$ m/s, leading to the maximum Doppler frequency shift of $f_{\max}^{\text{d}}=3000$ Hz. The total number of subcarriers is set to $N = 1024$, and the transmission numerology of $\mu=5$, leading to the subcarrier spacing of $\Delta f=480$ kHz, and the symbol interval of $T_{\text{s}} = 2.08$ us. 
In our simulation, we set $N_{\text{s}}=7$ symbols in one mini-slot, and the maximum phase shift under the maximum Doppler frequency shift is only 0.0438 rad. Hence, the channel stays constant in one mini-slot. The SNR is defined as $\mathrm{SNR}\triangleq \small{\frac{P\sigma _{\alpha}^{2}}{\sigma _{\text{n}}^{2}}}$, where $P$ denotes the average power of each entry in ${\bf S}$. In our simulations, the entries in ${\bf S}$ and ${\bf W}$ are randomly generated, and specifically, $\left\| \mathbf{S}_{:,n} \right\|^2=\frac{1}{N_{\text{MS}}}, n=1,2,\cdots,N_{\text{s}}$ to guarantee the power limit of hybrid precoding. 

\begin{figure}
	\centering
	\subfigure[MSE of AOAs]{
		\label{AOAMSE} 
		\includegraphics[width=0.48\linewidth]{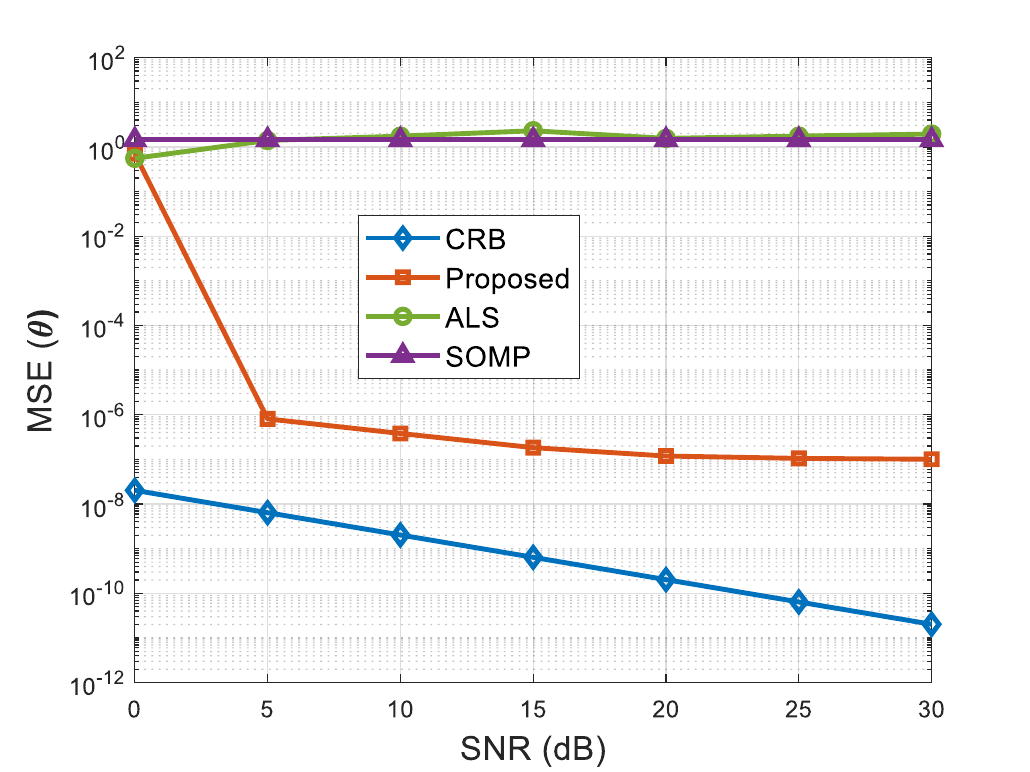}}
	\subfigure[MSE of AODs]{
		\label{AODMSE} 
		\includegraphics[width=0.48\linewidth]{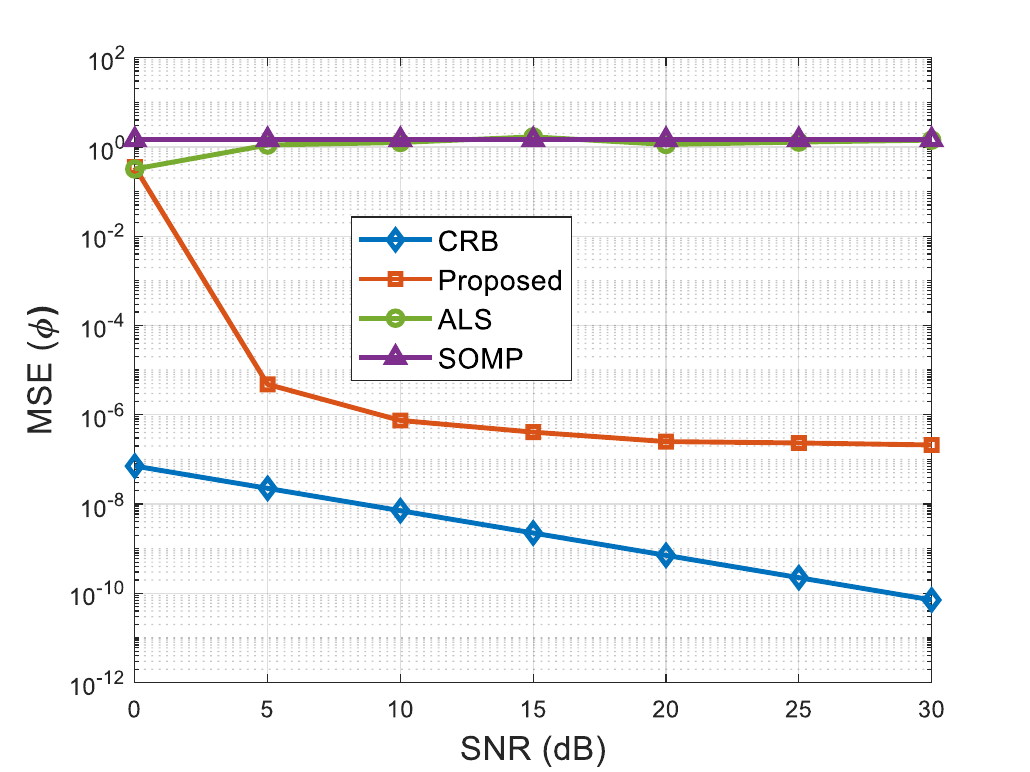}}
	\subfigure[MSE of Delays]{
		\label{delayMSE} 
		\includegraphics[width=0.48\linewidth]{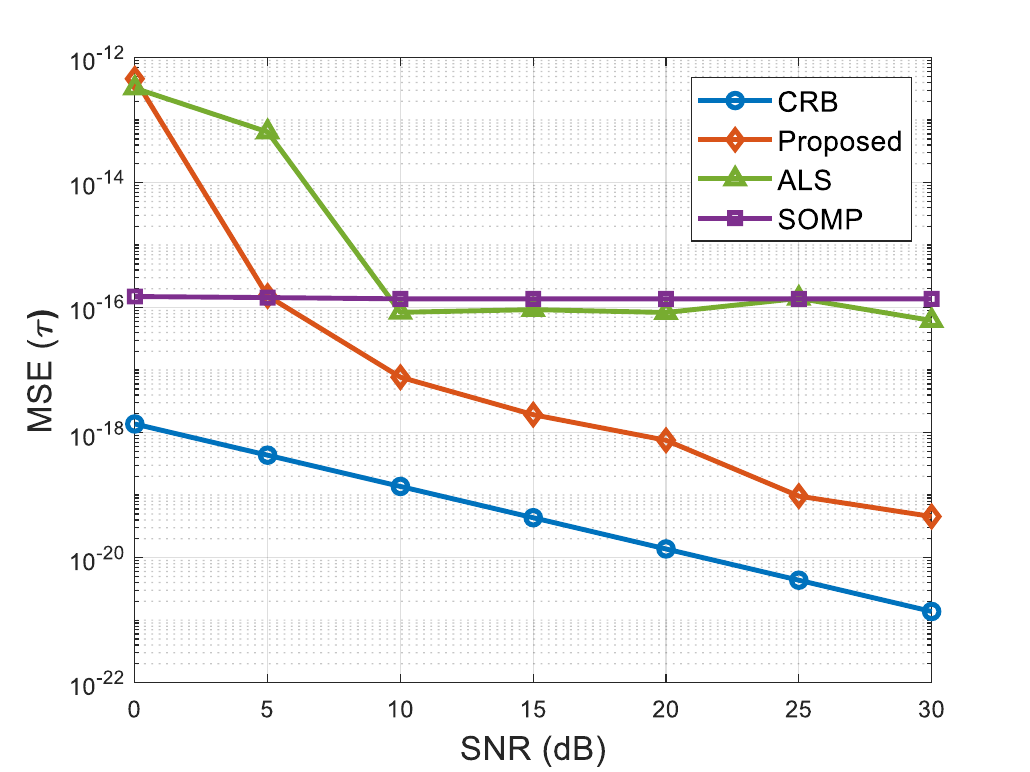}}
	\subfigure[MSE of Channel Gains]{
		\label{alphaMSE} 
		\includegraphics[width=0.48\linewidth]{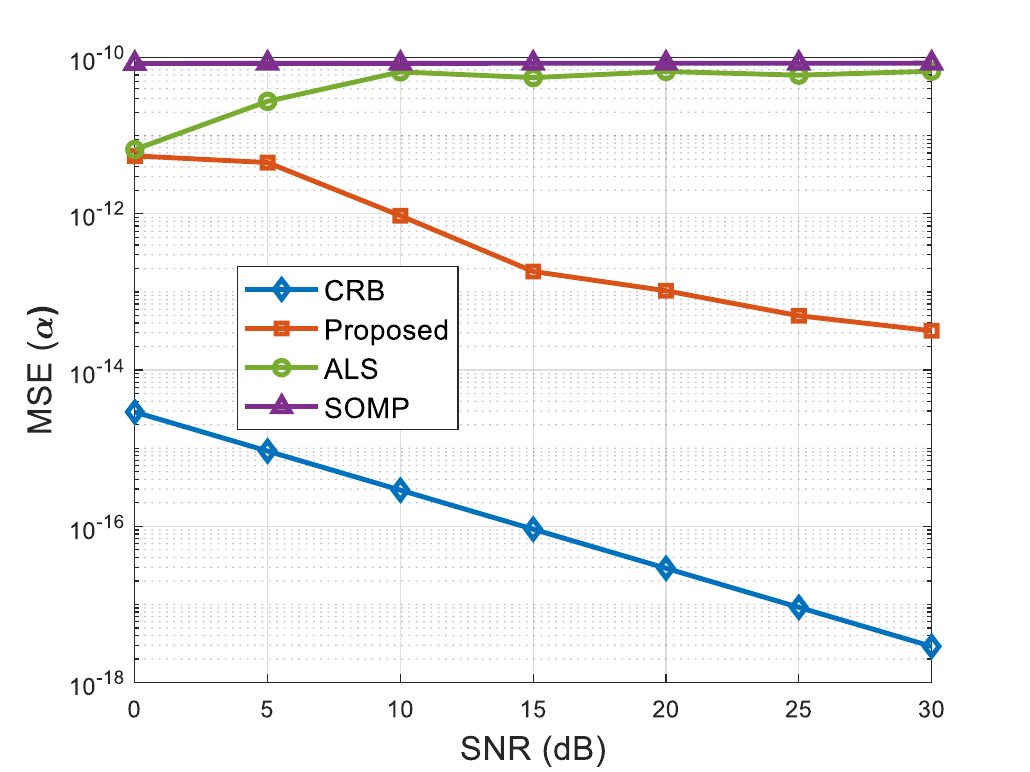}}
	\subfigure[MSE of Doppler Shifts]{
		\label{FdMSE} 
		\includegraphics[width=0.48\linewidth]{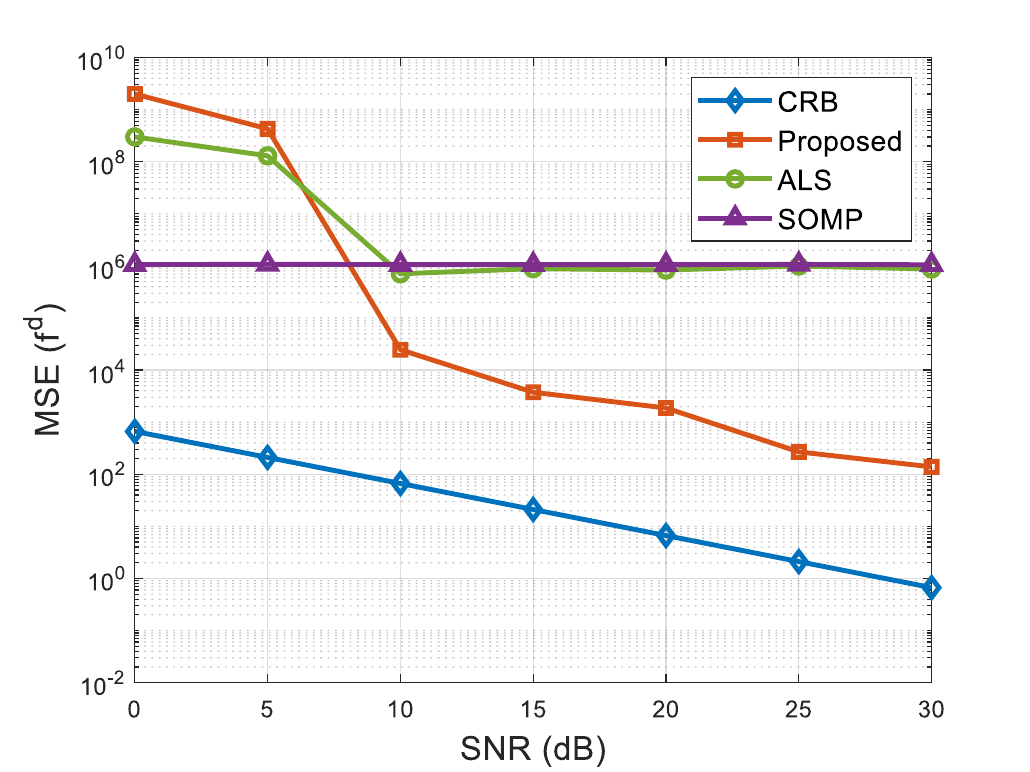}}
	\caption{MSE performance of the channel parameters versus SNR}
	\label{MSE} 
	\vspace{-0.7cm}
\end{figure}

 The estimation performance of the channel parameters are firstly examined. The corresponding mean square errors (MSEs) are computed for each set of channel parameters, which are given by
 
\begin{align}
	&\mathrm{MSE}\left( \mathrm{\theta} \right) =\left\| \bm{\theta }-\widehat{\bm{\theta}} \right\| ^{2},\,\, \mathrm{MSE}\left( \phi \right) \left\| \bm{\phi }-\widehat{\bm{\phi}} \right\| ^{2},\nonumber
	\\
	&\mathrm{MSE}\left( \tau \right) =\left\| \bm{\tau }-\widehat{\bm{\tau}} \right\| ^{2},\,\,   \mathrm{MSE}\left( \tau \right) =\left\| \bm{\alpha }-\widehat{\bm{\alpha}} \right\| ^{2},\nonumber
	\\
	&\mathrm{MSE}\left( f^{\text{d}} \right)= \left\| \mathbf{f}^{\text{d}}-\widehat{\mathbf{f}}^{\text{d}} \right\| ^{2},\nonumber
\end{align}
where $\bm{\theta }=\left[ \theta _1,\cdots ,\theta _L \right] ^{\mathrm{T}}$, $\bm{\phi }=\left[ \phi _1,\cdots ,\phi _L \right] ^{\mathrm{T}}$, $\bm{\tau }=\left[ \tau _1,\cdots ,\tau _L \right] ^{\mathrm{T}}$, $\bm{\alpha }=\left[ \alpha _1,\cdots ,\alpha _L \right] ^{\mathrm{T}}$ and ${\bf{f}}^{\text{d}}=\left[ f_{1}^{\text{d}},\cdots ,f_{L}^{\text{d}} \right] ^{\mathrm{T}}$, respectively. The ALS algorithm \cite{ryzhang,convergence2}, the SOMP algorithm \cite{{qinqibotvt,mmBOMP}} are selected as benchmarks. We also add the corresponding CRB results for different channel parameters for comparison. Fig.~\ref{MSE} depicts the MSE performance of the channel parameters of the proposed method versus the SNR. To approach the optimal solution of the maximum likelihood estimation problems (\ref{estAOA}) and (\ref{estAOD}), the angle search grid precision is set to $G=5000$. From Fig.~\ref{AOAMSE} and Fig.~\ref{AODMSE}, it can be observed that the estimations of AoAs and AoDs are very close to the corresponding CRBs between 5 dB and 15 dB. The performance becomes stable when the SNR is larger than 15 dB due to the limited search grid $G$. The MSE performance of delays is close to the CRB even when the SNR is low. The Doppler shift estimation is accurate when the SNR is larger than 15  dB. Finally, it can be seen from Fig.~\ref{alphaMSE} that the gap between the channel gain MSE and the CRB are wider than that of other channel parameters. This is due to the error propagation \cite{shijin} as the scaling ambiguities (\ref{delta1})-(\ref{delta3}) are first estimated and their estimate errors are accumulated in the estimation of the channel gain.

\begin{figure}[ht]
	\vspace{-0.2cm}
	\begin{minipage}[t]{1\linewidth}
		\centering
		\includegraphics[width=1.0\linewidth]{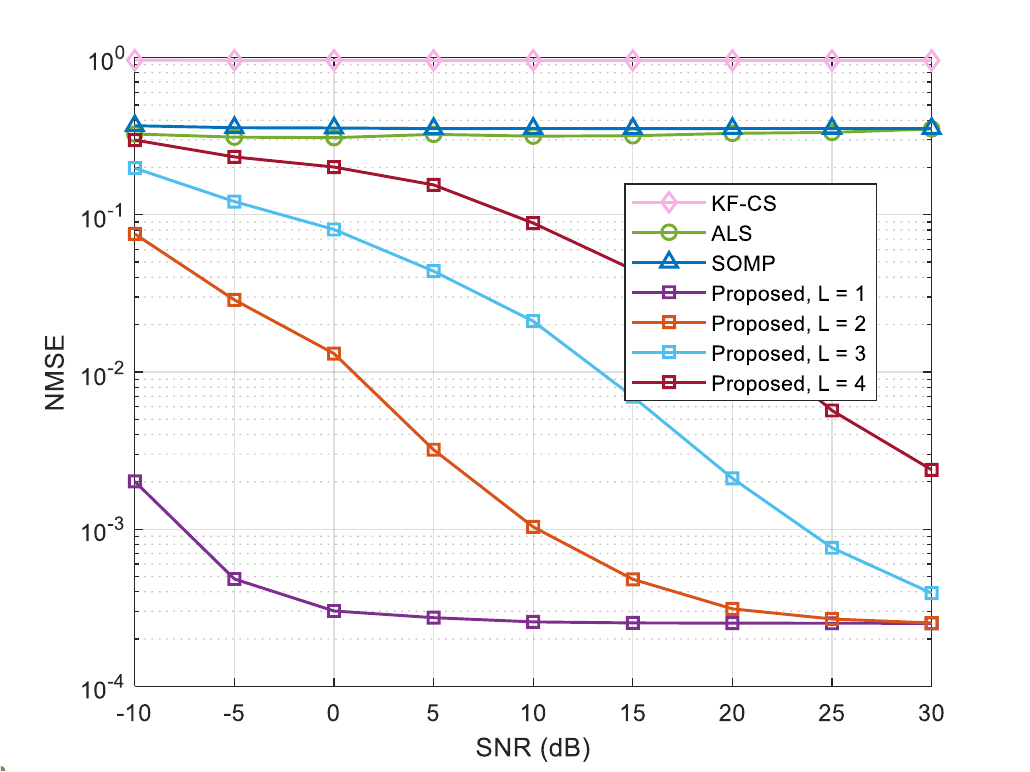}
		\vspace{-0.2cm}
		\caption{NMSE versus SNR.}
		\label{NMSEvsSNR}
	\end{minipage}%
	\hfill
	\vspace{-0.5cm}
\end{figure}

Then, we focus on the normalized mean square error (NMSE) performance of the channel estimation methods, with the focus on the comparison of the statistical channel model and the instantaneous channel model. The KF-CS method \cite{2024tcomSOMP}, the ALS-based method \cite{ryzhang,convergence2}, and the SOMP-based method \cite{qinqibotvt,mmBOMP} are selected as benchmarks. The KF-CS method is based on the statistical channel model, the ALS-based method and the SOMP-based method are based on the instantaneous channel model. The NMSE is defined as
\setcounter{equation}{62}
\begin{equation}
	\mathrm{NMSE}\ \small{\triangleq } \ \mathbb{E}\left\{\frac{1}{MK}\sum_{m=1}^M{\sum_{k=1}^K{\frac{\left\| \widehat{\mathbf{H}}_{m,k}-\mathbf{H}_{m,k} \right\| _{F}^{2}}{\left\| \mathbf{H}_{m,k} \right\| _{F}^{2}}}}\right\}.
\end{equation}
Fig.~\ref{NMSEvsSNR} shows the NMSE performance versus the SNR for the proposed method and benchmarks.  From Fig.~\ref{NMSEvsSNR}, it can be seen that the NMSE performance of the proposed method outperforms other algorithms. The main reasons are as follows. First, the KF-CS method is based on the statistical channel model, which is inaccurate to describe the phase time-variation in high-mobility scenarios compares to the instantaneous channel model. Second, the convergence of the ALS algorithm cannot be guaranteed as the rank of tensor (the number of channel paths) $L$ is larger than 2 \cite{convergence1,convergence2}. Compared with the ALS algorithm, the tensor decomposition algorithm does not need the iterations. Third, in the recovery of factor matrices ${\bf A}$ and ${\bf B}$, the columns ${\bf a}_l$ and ${\bf b}_l$ are obtained by performing SVD to $\widehat{\mathbf{E}}_l=\mathrm{unvec}_{Q_{\text{BS}}\times N_{\text s}}\left( \widehat{\mathbf{e}}_l \right) $, which is known as the optimum solution of the rank-one approximation Problem (\ref{AB}) \cite{SVD}, and the ML estimations are applied in angles estimation. Hence, the performance of angles estimation for tensor decomposition method is better than that of the SOMP-based method.

\begin{figure}[ht]
	\vspace{-0.2cm}
	\begin{minipage}[t]{1\linewidth}
		\centering
		\includegraphics[width=1.0\linewidth]{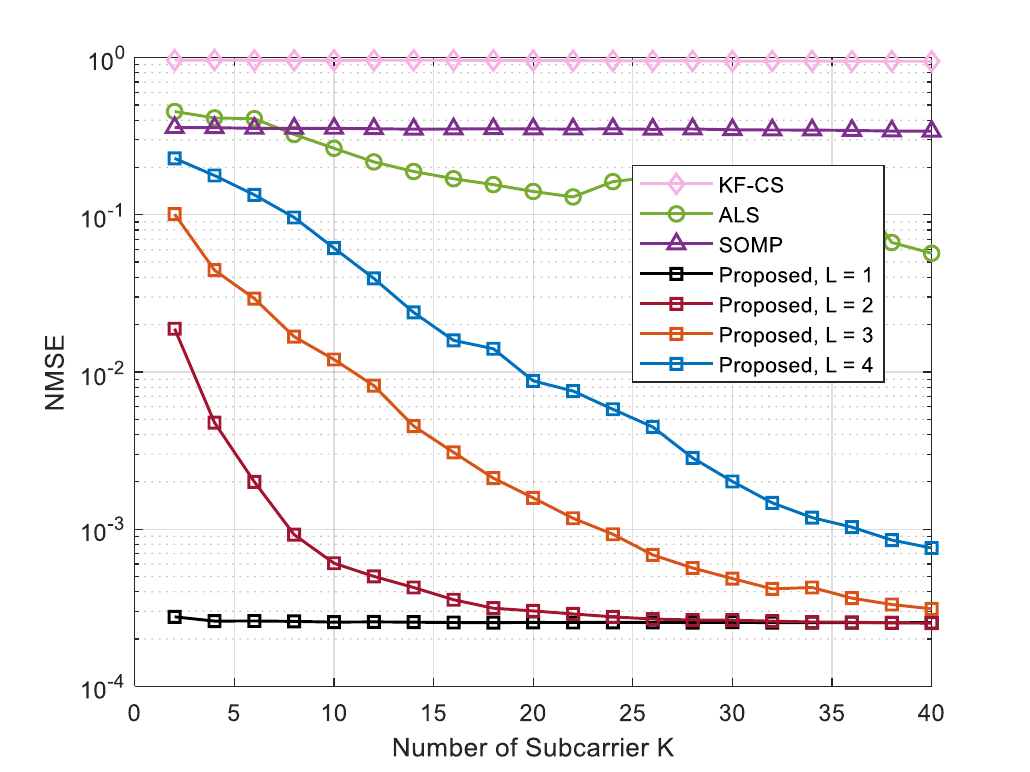}
		\vspace{-0.2cm}
		\caption{NMSE versus the number of subcarriers.}
		\label{NMSEvsK}
	\end{minipage}%
	\hfill
\end{figure}

In Fig.~\ref{NMSEvsK}, the NMSE performance versus the number of subcarriers is demonstrated, where the SNR is set to 10 dB. From Fig.~\ref{NMSEvsK}, it can be seen that the proposed algorithm outperforms the other methods.  When the number of channel paths is larger than one (e.g., $L>1$), the NMSE performance is not good when $K$ is small even though the uniqueness of multilinear decomposition matches. The channel estimation accuracy improved with the number of subcarriers $K$ increases for both tensor decomposition based channel estimation algorithms, but the similar trend does not present in the SOMP algorithm. For the proposed method, the number of subcarriers affects the estimation performance of the angles, delays and the channel gains. But for the SOMP algorithm, it only affects the delay estimation performance. Hence, the performance of the SOMP algorithm does not improve as the $K$ continuously increases. 

\begin{figure}[ht]
	\vspace{-0.2cm}
	\begin{minipage}[t]{1\linewidth}
		\centering
		\includegraphics[width=1.0\linewidth]{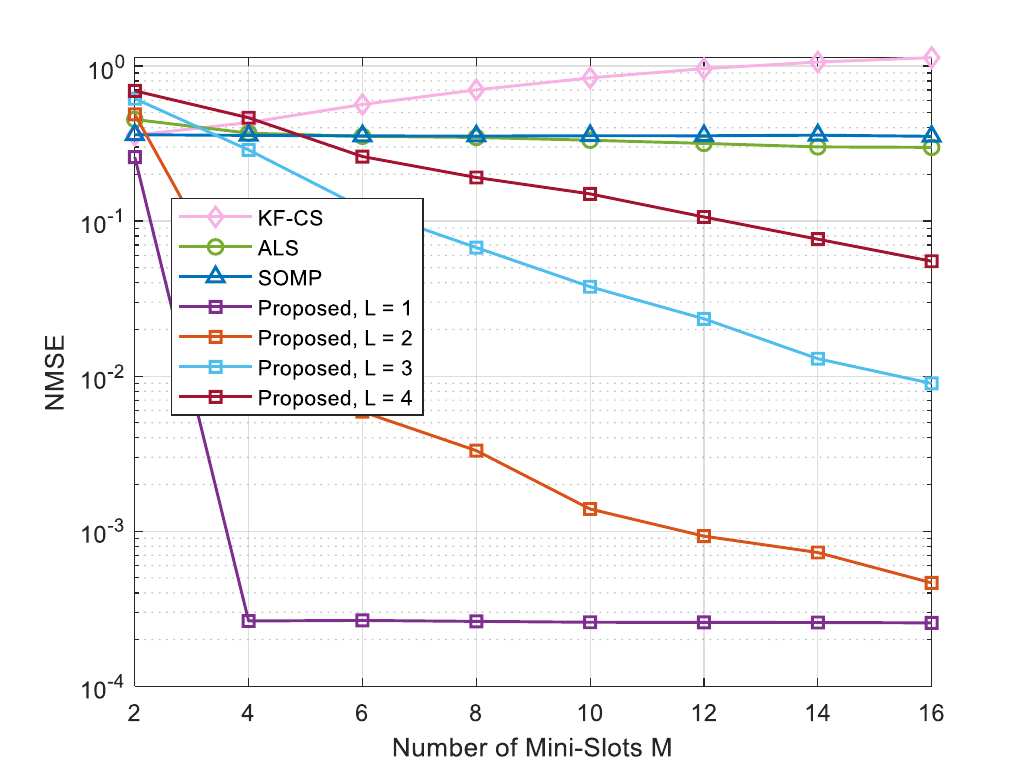}
		\vspace{-0.2cm}
		\caption{NMSE versus the number of mini-slots.}
		\label{NMSEvsM}
	\end{minipage}%
	\hfill
\end{figure}

The NMSE versus the number of mini-slots $M$ is shown in Fig.~\ref{NMSEvsM}, where the SNR is set to 10 dB. From Fig.~\ref{NMSEvsM}, the NMSE performance of the proposed method is not good when $M=2$, and significantly improves with the increase of $M$, which is similar to that in Fig.~\ref{NMSEvsK}. When $M$ increases, the estimate accuracy of the Doppler frequency shift effect $f_{l}^{\text{d}}$ improves, and the LS estimation (\ref{hatC}), (\ref{hatE}) can achieve better performance. Moreover, due to the assumption of inaccurate first-order AR model, the estimation error of the channel gains in previous slots will accumulate in the current slots, resulting in a decrease in the NMSE of the KF-CS method as $M$ increases.

\begin{figure}[ht]
	\vspace{-0.5cm}
	\begin{minipage}[t]{1\linewidth}
		\centering
		\includegraphics[width=1.0\linewidth]{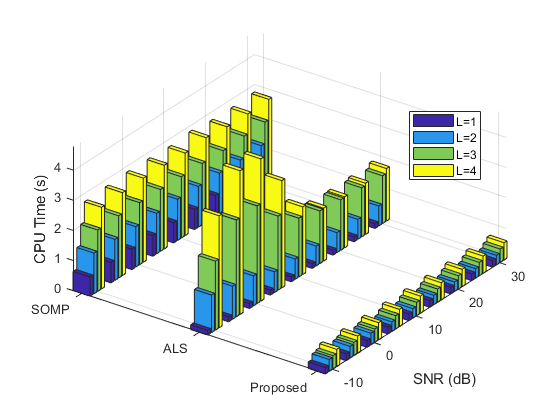}
		\vspace{-0.2cm}
		\caption{CPU Time versus SNR and the number of paths.}
		\label{cputime}
	\end{minipage}%
	\hfill
\end{figure}

Fig.~\ref{cputime} depicts the CPU time versus SNR and the number of paths for the proposed method, the ALS method and the SOMP method. From Fig.~\ref{cputime}, it can be seen that the CPU time of the proposed method and the SOMP method does not change with the SNR, but the CPU time for the ALS method is affected by the change of SNR significantly. The reason is that the number of iterations required in the ALS algorithm is larger when the SNR is low and medium. Furthermore, the CPU time significantly increases for the ALS method when the number of channel paths increases, as the ALS method needs more iterations to converge when the rank of tensor increases \cite{convergence1,convergence2}. However, the proposed method is still efficient even when the number of channel paths is $L=4$, as the proposed method does not need any iterations to recover the factor matrices.

\begin{figure}[ht]
	\vspace{-0.2cm}
	\begin{minipage}[t]{1\linewidth}
		\centering
		\includegraphics[width=1.0\linewidth]{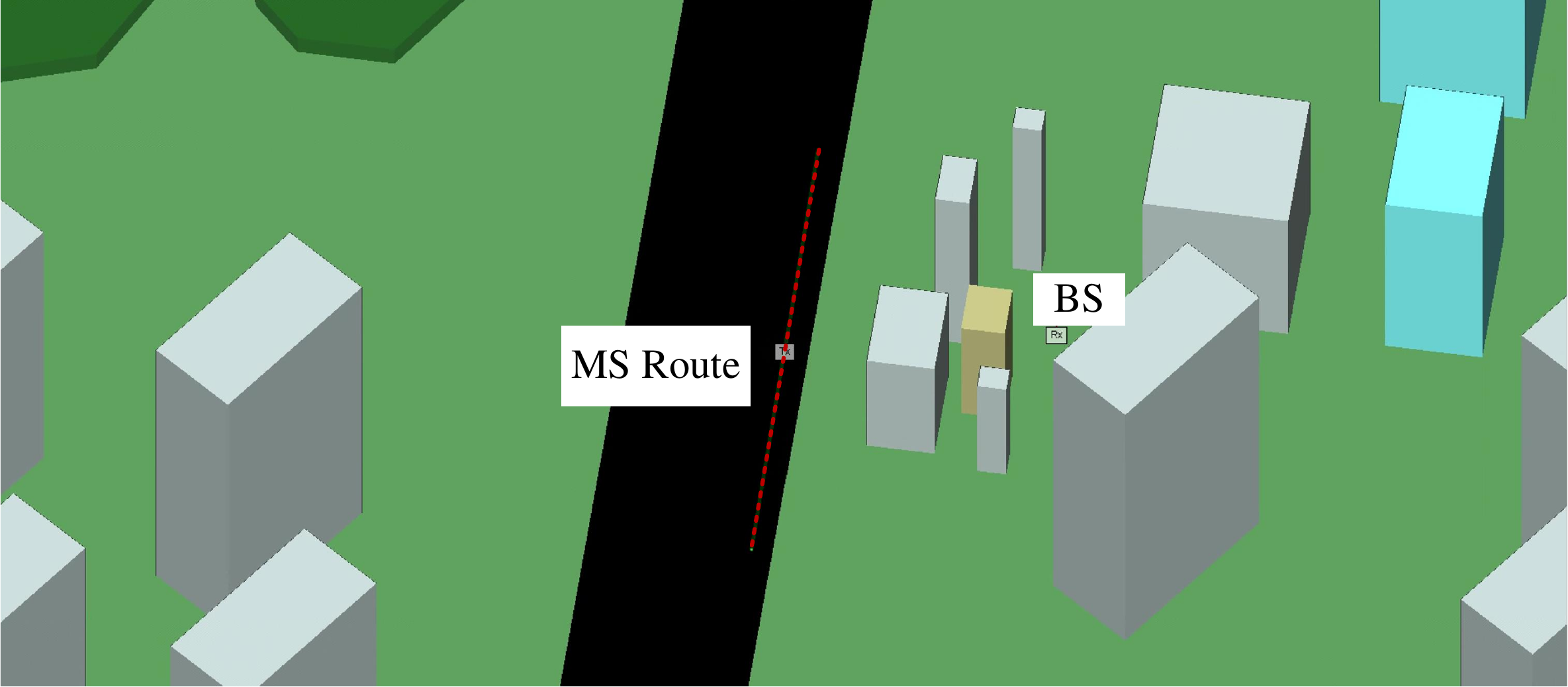}
		\vspace{-0.2cm}
		\caption{Wireless InSites Scenario.}
		\label{WI}
	\end{minipage}%
	\hfill
\end{figure}

\begin{figure}[ht]
	\vspace{-0.2cm}
	\begin{minipage}[t]{1\linewidth}
		\centering
		\includegraphics[width=1.0\linewidth]{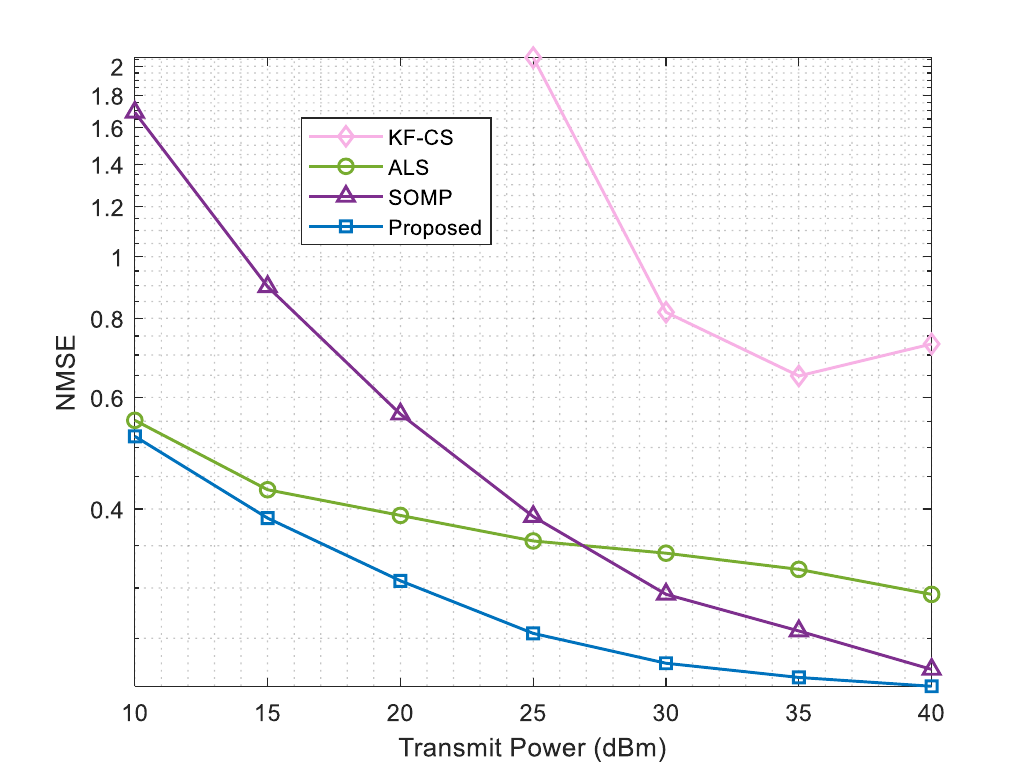}
		\vspace{-0.2cm}
		\caption{NMSE versus SNR, the channel parameters are generated from Wireless InSite.}
		\label{NMSEvsWI}
	\end{minipage}%
	\hfill
	\vspace{-0.5cm}
\end{figure}

To verify the channel estimation methods in more realistic and accurate channel samples, we use Wireless InSite \cite{WI} to generate the wireless environments. In Fig.~\ref{WI}, a highway MS-to-BS communication scenario is presented, where the buildings, forests, and other structures on both sides of the highway form the scatters of the wireless channel. In this scenario, the mobile station sends pilot signals to the static BS through uplink channel, and the BS estimates the channel based on the received signals. The number of antennas at the BS and that at the MS are set to 64 and 32, respectively. The speed of the MS is set to 30 m/s. The other parameters are the same as before. The Doppler shift for the $l$-th channel path is computed as \cite{WI}
\begin{equation}
	f^{\text{d}}_l=f_{\text c} \frac{\mathbf{d}_l\cdot \mathbf{v}_{\mathrm T}}{c}, \nonumber
\end{equation} 
where $\mathbf{v}_{\mathrm T}$ is the velocity vector of the transmitter, $\mathbf{d}_l$ is the directions of the departure of the $l$-th channel path, and $f_{\text c}$ is the carrier frequency, which is set to 30 GHz. In Fig.~\ref{NMSEvsWI}, the NMSE performance versus the SNR is depicted, where the channel parameters are generated from Wireless InSite scenario in Fig.~\ref{WI}. Due to the rich scatters in the scenario, the estimated number of paths is set to $L=4$. From Fig.~\ref{NMSEvsWI}, it can be seen that the proposed method outperforms the KF-CS method, the ALS method and the SOMP method. Note that the KF-CS method performs poorly in Wireless Insites scenario, indicating that the first-order AR model cannot accurately depict the time variant of the channel in high-mobility scenarios. Furthermore, even in scenarios with a large number of scatters, the proposed method still achieves acceptable performance, which demonstrates the effectiveness of the proposed method for channel estimation in practical scenarios.

\section{Conclusion}\label{conclusion}
In this paper, we proposed an ESPRIT-type decomposition based method for channel estimation for mmWave MIMO-OFDM systems in high-mobility scenarios. By stacking the pilot signals from space, time and frequency domains, a fourth-order tensor was constructed, which satisfies the low-rank CP model. Then, the channel estimation was formulated as a CP decomposition problem plus parameter estimation problems. To solve the CP decomposition problem, we utilized the property of Vandermonde matrix and ESPRIT based method to recover the factor matrices. The channel parameters, including angles, delays, channel gains and Doppler shifts, were estimated based on the recovered factor matrices. The following conclusions can be drawn from the simulation results. First, the results reveal that the statistical channel model (i.e., the first-order AR model) is inaccurate in high-mobility scenarios, and the Doppler frequency shift estimation should be considered. Second, due to the fact that the tensor decomposition based methods first decompose the received signals into factor matrices, where each column of the factor matrix corresponds to a channel path, and then estimates the parameters, the impact caused by non-orthogonality between different paths in the channel can be reduced, thus achieving better performance than CS-based methods. Third, the proposed method is based on the spatial smoothing and ESPRIT algorithm, which can achieve higher efficiency in scenarios with rich scatters than the ALS-based method and the CS-based methods. Finally, simulation results based on the scenario generated by Wireless Insites reveals the inaccurate of statistical channel model and verified the effectiveness of the proposed algorithm based on the instantaneous channel model in practical scenarios.
\vspace{-0.3cm}

\begin{appendix}
The mode-1 unfolding of $\mathcal{Z}$ is  $\mathbf{Z}_1=\left( \mathbf{D}\odot \mathbf{C}\odot \mathbf{B} \right) \mathbf{A}^{\mathrm{T}}$. Due to the property of the Khatri-Rao product $\mathrm{vec}\left( \mathbf{UVW} \right) =\left( \mathbf{W}^{\mathrm{T}}\odot \mathbf{U} \right) \text{d}\left( \mathbf{V} \right) $, where ${\bf V}$ is a diagonal matrix, we have
\begin{align}
	\mathrm{vec}\left( \mathbf{Z}_1 \right) =\left( \mathbf{A}\odot \mathbf{D}\odot \mathbf{C}\odot \mathbf{B} \right) \mathbf{1}_L
	\nonumber\\
	=\sum_{l=1}^L{\mathbf{a}_l\otimes \mathbf{d}_l\otimes \mathbf{c}_l\otimes \mathbf{b}_l}.
\end{align}

Next, the partial derivatives of $L({\bf p})$ w.r.t. channel parameters ${\bf p}$ are derived. The partial derivative of $L({\bf p})$ to $\theta_l$ is given by
\begin{equation}\label{Lptheta}
	\frac{\partial L\left( \mathbf{p} \right)}{\partial \theta _l}=\left( \frac{\partial L\left( \mathbf{p} \right)}{\partial \mathbf{z}_1} \right) ^{\mathrm T}\frac{\partial \mathbf{z}_1}{\partial \theta _l}+\left( \frac{\partial L\left( \mathbf{p} \right)}{\partial \mathbf{z}_{1}^{\ast}} \right) ^{\mathrm T}\frac{\partial \mathbf{z}^{\ast}_1}{\partial \theta _l}, 
\end{equation}
where the partial derivative to $L({\bf p})$ w.r.t. ${\bf z}_1$ is given by
\begin{align}\label{Lp/z}
\frac{\partial L\left( \mathbf{p} \right)}{\partial \mathbf{z}_1}&=\left( \mathbf{C}_{\mathcal{N}} \right) ^{-\mathrm{T}}\left( \mathbf{y}_1-\mathbf{z}_1 \right) ^{\ast}
\nonumber\\
&=\left( \mathbf{C}_{\mathcal{N}} \right) ^{-\mathrm{T}}\mathrm{vec}\left( \mathbf{N}_{\left( 1 \right)}^{\mathbf{W}} \right) ^{\ast}.
\end{align}
Due to the fact that only the fiber ${\bf a}_l$ is associated with $\theta_l$, the partial derivative to ${\bf z}_1$ w.r.t. $\theta_l$ is give by
\begin{align}
\frac{\partial \mathbf{z}_1}{\partial \theta _l}&=\frac{\partial \left( \sum_{r=1}^L{\mathbf{a}_r\otimes \mathbf{d}_r\otimes \mathbf{c}_r\otimes \mathbf{b}_r} \right)}{\partial \theta _l}\nonumber\\
&=\left( \frac{\partial \mathbf{a}_l}{\partial \theta _l} \right) \otimes \mathbf{d}_l\otimes \mathbf{c}_l\otimes \mathbf{b}_l, \label{z/theta}
\end{align}
where 
\begin{align}\label{al/thetal}
	\frac{\partial \mathbf{a}_l}{\partial \theta _l}&=\mathbf{W}^{\mathrm{T}}\left[
	0,-j2\pi \frac{d}{\lambda}\sin \theta _le^{j2\pi \frac{d}{\lambda}\cos \theta _l},\right. \nonumber\\
	&\qquad\left.\cdots, -j2\pi \frac{d}{\lambda}\left( N_{\text {BS}}-1 \right) \sin \theta _le^{j2\pi \frac{d}{\lambda}\left( N_{\text {BS}}-1 \right) \cos \theta _l}\right] ^{\mathrm{T}}.
\end{align}
Substituting (\ref{Lp/z})-(\ref{al/thetal}) into (\ref{Lptheta}), we have 
\begin{equation}\label{realLptheta}
	\frac{\partial L\left( \mathbf{p} \right)}{\partial \theta _l}=	2\mathrm{Re}\left\{ \mathrm{vec}\left( \mathbf{N}_{\left( 1 \right)}^{\mathbf{W}} \right) ^{\mathrm{H}}\left( \mathbf{C}_{\mathcal{N}} \right) ^{-1}\frac{\partial \mathbf{z}_1}{\partial \theta _l} \right\}.
\end{equation}
The partial derivation of $L({\bf p})$ w.r.t. other parameters can be similarly derived according to (\ref{Lptheta})-(\ref{realLptheta}),
where the partial derivatives of $L({\bf p})$ w.r.t. the corresponding channel parameters are given by
\begin{align}
\frac{\partial \mathbf{z}_1}{\partial \phi _l}&=\frac{\partial \sum_{l=1}^L{{\mathbf{a}}_l\otimes {\mathbf{d}}_l\otimes {\mathbf{c}}_l\otimes {\mathbf{b}}_l}}{\partial \phi _l}\nonumber\\
&={\mathbf{a}}_l\otimes {\mathbf{d}}_l\otimes {\mathbf{c}}_l\otimes \left( \frac{\partial {\mathbf{b}}_l}{\partial \phi _l} \right)\nonumber
\end{align} 
\begin{align}
\frac{\partial \mathbf{z}_1}{\partial \tau _l}&=\frac{\partial \sum_{l=1}^L{{\mathbf{a}}_l\otimes {\mathbf{d}}_l\otimes {\mathbf{c}}_l\otimes {\mathbf{b}}_l}}{\partial \tau _l}\nonumber\\
&={\mathbf{a}}_l\otimes {\mathbf{d}}_l\otimes \left( \frac{\partial {\mathbf{c}}_l}{\partial \tau _l} \right) \otimes {\mathbf{b}}_l\nonumber
\end{align}
\begin{align}
\frac{\partial \mathbf{z}_1}{\partial \alpha _l}&=\frac{\partial \sum_{l=1}^L{{\mathbf{a}}_l\otimes {\mathbf{d}}_l\otimes {\mathbf{c}}_l\otimes {\mathbf{b}}_l}}{\partial \alpha _l}\nonumber\\
&={\mathbf{a}}_l\otimes {\mathbf{d}}_l\otimes \left( \frac{\partial {\mathbf{c}}_l}{\partial \alpha _l} \right) \otimes {\mathbf{b}}_l\nonumber
\end{align}
\begin{align}
\frac{\partial \mathbf{z}_1}{\partial f^{\text d}_{l}}&=\frac{\partial \sum_{l=1}^L{{\mathbf{a}}_l\otimes {\mathbf{d}}_l\otimes {\mathbf{c}}_l\otimes {\mathbf{b}}_l}}{\partial f^{\text d}_{l}}\nonumber\\
&\overset{a}{=}{\mathbf{a}}_l\otimes \frac{\partial \tilde{\mathbf{d}}_l}{\partial f^{\text d}_{l}}\otimes {\mathbf{c}}_l\otimes {\mathbf{b}}_l\nonumber\\
&\qquad+j2\pi \tau _le^{j2\pi f^{\text d}_{l}\tau _l}{\mathbf{a}}_l\otimes {\mathbf{d}}_l\otimes {\mathbf{c}}_l\otimes {\mathbf{b}}_l,\nonumber
\end{align}
where equation $a$ is due to the fact that both the fiber ${\bf c}_l$ and the fiber ${\bf d}_l$ are associated with the Doppler shift $f^{\text{d}}_l$. The partial derivatives to the fibers w.r.t. the corresponding channel parameters are given by
\begin{align}
\frac{\partial {\mathbf{b}}_l}{\partial \phi _l}&=\mathbf{C}^{\mathrm{T}}\left[ 0,-j2\pi \frac{d}{\lambda}\sin \phi _le^{j2\pi \frac{d}{\lambda}\cos \phi _l},\right.\nonumber\\
&\qquad \left.\cdots ,-j2\pi \frac{d}{\lambda}\left( N_{\text{MS}}-1 \right) \sin \phi _le^{j2\pi \frac{d}{\lambda}\left( N_{\text{MS}}-1 \right) \cos \phi _l} \right] ^{\mathrm{T}}\nonumber
\\
\frac{\partial {\mathbf{c}}_l}{\partial \tau _l}&=\alpha _l\left[ -j\frac{2\pi}{N}\left( f_{\text s}-Nf_{l}^{\text{d}} \right) e^{-j\frac{2\pi}{N}\left( f_{\text s}\tau _l \right) \left( 1-\frac{f_{l}^{\text{d}}}{f_{\text s}}N \right)},\right.\nonumber\\
&\qquad\left.\cdots ,-j\frac{2\pi}{N}\left( Kf_{\text s}-Nf_{l}^{\text{d}} \right) e^{-j\frac{2\pi}{N}\left( f_{\text s}\tau _l \right) \left( K-\frac{f_{l}^{\text{d}}}{f_{\text s}}N \right)} \right] ^{\mathrm{T}}\nonumber
\\
\frac{\partial {\mathbf{c}}_l}{\partial \alpha _l}&=\left[ e^{-j\frac{2\pi}{N}\left( f_{\text s}\tau _l \right) \left( 1-\frac{f_{l}^{\text{d}}}{f_{\text s}}N \right)},\cdots ,e^{-j\frac{2\pi}{N}\left( f_{\text s}\tau _l \right) \left( K-\frac{f_{l}^{\text{d}}}{f_{\text s}}N \right)} \right] ^{\mathrm{T}}\nonumber
\\
\frac{\partial {\mathbf{d}}_l}{\partial f_{l}^{\text{d}}}&=\left[ 0,j2\pi N_{\text s}T_{\text s}e^{j2\pi f_{l}^{\text{d}}N_{\text s}T_{\text s}},\right.\nonumber\\
&\qquad\left.\cdots ,j2\pi \left( M-1 \right) N_{\text s}T_{\text s}e^{j2\pi f_{l}^{\text{d}}\left( M-1 \right) N_{\text s}T_{\text s}} \right] ^{\mathrm{T}}.
\end{align}
According to the above discussion, we can derive the FIM $\mathbf{I}\left( {\bf p} \right) =\mathbb{E} \left\{ \left( \frac{\partial L\left( \mathbf{p} \right)}{\partial \mathbf{p}} \right) ^{\mathrm H}\left( \frac{\partial L\left( \mathbf{p} \right)}{\partial \mathbf{p}} \right) \right\} $. We first introduce the derivation of the special case $\mathbb{E} \left\{ \left( \frac{\partial L\left( \mathbf{p} \right)}{\partial \theta _i} \right) ^{\ast}\frac{\partial L\left( \mathbf{p} \right)}{\partial \theta _j} \right\} $, and the others can be similarly derived. 

Based on (\ref{Lptheta})-(\ref{realLptheta}), we can obtain
\begin{align}\label{crbtheta}
	&\mathbb{E} \left\{ \left( \frac{\partial L\left( \mathbf{p} \right)}{\partial \theta _i} \right) ^{\ast}\frac{\partial L\left( \mathbf{p} \right)}{\partial \theta _j} \right\} \nonumber\\
	&\quad\overset{a}{=}\left( \frac{\partial \mathbf{z}_1}{\partial \theta _i} \right) ^{\mathrm{T}}\mathbf{C}_{\mathcal{N}}^{-1}\left( \frac{\partial \mathbf{z}_1}{\partial \theta _j} \right) ^{\ast}+\left( \frac{\partial \mathbf{z}_1}{\partial \theta _i} \right) ^{\mathrm{H}}\mathbf{C}_{\mathcal{N}}^{-\mathrm{T}}\frac{\partial \mathbf{z}_1}{\partial \theta _j}\nonumber\\
	&\quad=	2\left( \frac{\partial \mathbf{z}_1}{\partial \theta _i} \right) ^{\mathrm{T}}\mathbf{C}_{\mathcal{N}}^{-1}\left( \frac{\partial \mathbf{z}_1}{\partial \theta _j} \right) ^{\ast},
\end{align}
where the equation $a$ is due to the fact that the entries in $\mathcal{N}$ follow the i.i.d circular symmetric complex Gaussian distribution, and thus the 2-order moment satisfies \cite{zhouzhoujsac}
\begin{align}
	&\mathbb{E} \left\{ \mathrm{vec}\left( \mathbf{N}_{\left( 1 \right)}^{\mathbf{W}} \right) \mathrm{vec}\left( \mathbf{N}_{\left( 1 \right)}^{\mathbf{W}} \right) ^{\mathrm{T}} \right\} \nonumber\\
	&=\mathbb{E} \left\{ \mathrm{vec}\left( \mathbf{N}_{\left( 1 \right)}^{\mathbf{W}} \right) ^{\ast}\mathrm{vec}\left( \mathbf{N}_{\left( 1 \right)}^{\mathbf{W}} \right) ^{\mathrm{H}} \right\}\nonumber\\
	&=\mathbf{0}_{Q_{\text{BS}}N_{\text s}MK\times Q_{\text{BS}}N_{\text s}MK}.\nonumber
\end{align}

The other elements in the FIM ${\bf I}({\bf p})$ can be computed similarly    to (\ref{crbtheta}). Without loss of generality, define
\begin{align}
	\mathbf{D}_{\mathbf{p}}\triangleq \frac{\partial \mathbf{z}_1}{\partial \mathbf{p}^T}=\left[ \frac{\partial \mathbf{z}_1}{\partial \theta _1},		\cdots,\frac{\partial \mathbf{z}_1}{\partial \theta _L},		\cdots,\frac{\partial \mathbf{z}_1}{\partial f_{L}^{\text{d}}}\right], 
\end{align} 
and the $(i,j)$-th entry is given by
\begin{equation}
	\left[ \mathbf{I}\left( \mathbf{p} \right) \right] _{i,j}=2\left( \mathbf{D}_{\mathbf{p}} \right)^{\mathrm{T}} _{:,i}\mathbf{C}_{\mathcal{N}}^{-1}\left( \mathbf{D}_{\mathbf{p}}^{\ast} \right) _{:,j}.
\end{equation}

\end{appendix}

\bibliographystyle{IEEEtran}
\bibliography{myre}


\end{document}